\documentclass[fleqn,10pt]{wlscirep}
\usepackage[title]{appendix}

\usepackage[utf8]{inputenc}
\usepackage[T1]{fontenc}
\usepackage{microtype}
\usepackage{graphicx}
\usepackage[export]{adjustbox}
\usepackage{subfigure}
\usepackage{booktabs} 
\usepackage{hyperref}

\RequirePackage{algorithm,algpseudocode}
\usepackage{amsmath}
\usepackage{amssymb}
\usepackage{mathtools}
\usepackage{amsthm}
\usepackage[capitalize,noabbrev]{cleveref}

\usepackage[textsize=tiny]{todonotes}
\title{Estimating the number of communities in weighted networks}
\author[*]{Huan Qing}
\affil[*]{China University of Mining and Technology, School of Mathematics, Xuzhou, 221116, P.R. China}
\affil[*]{qinghuan@cumt.edu.cn;qinghuan07131995@163.com}
\keywords{Community detection, distribution-free model, spectral clustering, weighted modularity, weighted networks}

\begin{abstract}
Community detection in weighted networks has been a popular topic in recent years. However, while there exist several flexible methods for estimating communities in weighted networks, these methods usually assume that the number of communities is known. It is usually unclear how to determine the exact number of communities one should use. Here, to estimate the number of communities for weighted networks generated from arbitrary distribution under the degree-corrected distribution-free model, we propose one approach that combines weighted modularity with spectral clustering. This approach allows a weighted network to have negative edge weights and it also works for signed networks. We compare the proposed method to several existing methods and show that our method is more accurate for estimating the number of communities both numerically and empirically.
\end{abstract}
\begin{document}

\flushbottom
\maketitle
%
%
\thispagestyle{empty}
\section{Introduction}\label{sec1}
For decades, network science provided substantial quantitative tools for the study of complex systems \cite{barabasi1999emergence,albert2002statistical,newman2003structure,boccaletti2006complex}. Networks emerge in numerous fields including physics, sociology, biology, economics, and so forth \cite{lusseau2004identifying,guimera2005functional,barabasi2004network,palla2007quantifying, bullmore2009complex,foster2005simplistic,schweitzer2009economic, pastor2015epidemic}. The elementary parts of a network are nodes, links, and link weights. A network is unweighted when all link weights are 1 and weighted otherwise \cite{newman2004analysis}. Networks usually have community structure such that nodes within the same community have more connections than across communities \cite{fortunato2010community,fortunato2016community}. For example, in social networks, communities can be groups of students who belong to the same school, be of the same club, be of the same graduation year, or be interested in the same movie; in scientific collaboration networks, communities are scientists in the same field \cite{newman2001structure,ji2016coauthorship,ji2022co}; in protein-protein interaction networks, communities are proteins enjoying similar functions \cite{schwikowski2000network,ideker2008protein}. However, in practice, the latent community structure of a network is generally not directly observable and we need to develop techniques to infer community structure.

Community detection for unweighted networks has been widely studied for decades \cite{fortunato2010community,fortunato2016community}. Numerous community detection methods are developed to fit a statistical model that can generate a random network with a community structure. The stochastic blockmodels (SBM) \cite{SBM} is a classical and popular generative model for unweighted networks. The popular degree-corrected stochastic blockmodels (DCSBM) extends SBM by considering node heterogeneity. Based on SBM and DCSBM, substantial community detection methods have been developed, such as \cite{rohe2011spectral,amini2013pseudo, RSC,lei2015consistency, SCORE,joseph2016impact, GeoNMF,chen2018convexified, OCCAM, MaoSVM,mao2020estimating,li2021convex,jing2022community}. However, most community detection methods require that the number of communities $K$ should be known in advance, and this is often not the case for real-world unweighted networks. To address this problem, some methods with theoretical guarantees are developed to estimate $K$ under SBM or DCSBM \cite{newman2016estimating,bickel2016hypothesis,lei2016goodness,riolo2017efficient,saldana2017how,wang2017likelihood,yan2018provable,chen2018network,ma2021determining,le2022estimating}, where the spectral clustering methods developed in \cite{le2022estimating} stand out as they estimate $K$ for unweighted networks regardless of statistical models.

A significant drawback of the above SBM-based and DCSBM-based methods is that they ignore the impact of edge weights, i.e., they only consider unweighted networks and ignore weighted networks. Edge weights are common in network data and they could help us to understand the community structure of a network better \cite{newman2004analysis}. In recent years, community detection in weighted networks is a hot topic and many statistical models have been developed to fit weighted networks, such as the weighted stochastic blockmodels (WSBM) proposed in \cite{aicher2015learning, jog2015information,ahn2018hypergraph, palowitch2018significance,peixoto2018nonparametric,xu2020optimal,ng2021weighted}, the distribution-free model (DFM) of \cite{DFM}, and the degree-corrected distribution-free model (DCDFM) introduced in \cite{DCDFM}. Among these models for weighted networks, the DFM model and its extension DCDFM stand out as they allow edge weights to follow any distribution as long as the expected adjacency matrix follows a block structure related to community partition. However, similar to SBM-based and DCSBM-based methods, algorithms developed for the above models modeling weighted networks also assume that the number of communities $K$ is known in advance, which is usually impractical for real-world weighted networks. To close this gap, we provide a simple approach to estimate $K$ for weighted networks generated from DCDFM.

The main contributions of this work include:

(1) We propose a simple method by taking advantage of both spectral clustering and weighted modularity to estimate the number of communities for weighted networks. The method determines $K$ by increasing the number of communities until weighted modularity does not increase. The method can estimate the number of communities of weighted networks generated from arbitrary distribution under DCDFM. The method is devised for DCDFM, but it can be naturally applied to weighted networks generated from DFM and unweighted networks generated from SBM and DCSBM since these three models are sub-models of DCDFM.

(2) We conduct a large number of experiments on both computer-generated weighted networks and real-world networks including signed networks. The experimental results show that our method proposed in this paper can estimate the number of communities when the weighted network is generated by different distributions under DCDFM even when the true $K$ is 1 and it is more accurate than its competitors.
\section{Methodology}
\subsection{The degree-corrected distribution-free model}
In this article, we work with the degree-corrected distribution-free model proposed in \cite{DCDFM}. We assume that there exist $K$ perceivable non-overlapping clusters
$\mathcal{C}^{(1)}, \mathcal{C}^{(2)}, \ldots, \mathcal{C}^{(K)}$, and each node only belongs to exactly one cluster. Let the $n\times 1$ vector $\ell$ denote the node label such that $\ell_{i}$ takes value from $\{1,2,\ldots, K\}$ and $\ell_{i}$ is the community label for node $i$ for $i\in[n]$. Let $Z\in\{0,1\}^{n\times K}$ be the community membership matrix such that $Z_{ik}=1$ if $\ell_{i}=k$ and $Z_{ik}=0$ otherwise. Let $\theta$ be an $n\times 1$ vector such that the positive number $\theta_{i}$ is the node heterogeneity of node $i$. Let $\Theta$ be an $n\times n$ diagonal matrix whose $i$-th diagonal entry is $\theta_{i}$. Let $P$ be the $K\times K$ symmetric connectivity matrix such that $P$'s rank is $K$, $P$'s elements can be any real values in $[-1,1]$, and $\mathrm{max}_{k,l\in[K]}|P_{kl}|=1$, where we let $P$'s maximum absolute element be 1 for convenience since we consider the node heterogeneity parameter $\theta$. For $i,j\in[n]$, the DCDFM model \cite{DCDFM} generates the  $(i,j)$-th element of the symmetric adjacency matrix $A$ for an un-directed weighted network $\mathcal{N}$ in the following way:
\begin{align}\label{ADCDFM}
\Omega:=\Theta ZPZ'\Theta~~~A_{ij}\mathrm{~is~a~random~variable~generated~from~arbitrary~distribution~}\mathcal{F}\mathrm{~with~expectation~}\Omega_{ij}.
\end{align}

DCDFM includes several previous models. For example, when $\theta_{i}=\sqrt{\rho}$ for all $i\in[n]$, DCDFM reduces to the distribution-free model \cite{DFM}; when $\mathcal{F}$ is Bernoulli distribution and $P$'s elements are nonnegative, DCDFM reduces to the classical degree-corrected stochastic blockmodels \cite{DCSBM}; when $\mathcal{\mathcal{F}}$ is Bernoulli distribution, all elements of $\theta$ are the same, and $P$'s elements are nonnegative, DCDFM reduces to the popular stochastic blockmodels \cite{SBM}, i.e., SBM, DCSBM, and DFM are sub-models of DCDFM. As analyzed in \cite{DCDFM}, $\mathcal{F}$ can be any distribution as long as $A$'s expectation matrix is $\Omega$ under distribution $\mathcal{F}$. Meanwhile, the fact that whether $P$'s elements can be negative depends on distribution $\mathcal{F}$. For example, when $\mathcal{F}$ is Bernoulli, Binomial, Poisson, Geometric or Exponential distributions, $P$'s elements should be nonnegative or positive; when $\mathcal{F}$ is Normal, Laplace or $A$ is the adjacency matrix of a signed network, $P$'s elements can be negative. DCDFM can generate $A$ for weighted networks benefitting from the arbitrariness of distribution $\mathcal{F}$.

When $n, K, \ell, P$, and $\theta$ are set, we can generate the adjacency matrix $A$ for any distribution $\mathcal{F}$ under DCDFM as long as Equation (\ref{ADCDFM}) holds. Given $A$ and the known number of clusters $K$, \cite{DCDFM} designs an efficient spectral algorithm called nDFA to estimate the node label vector $\ell$ and shows that nDFA enjoys consistent estimation under DCDFM for any distribution $\mathcal{F}$ satisfying Equation (\ref{ADCDFM}). However, the method nDFA requires $K$ to be known in advance, and this is not the case in practice.  To process this problem, in this article, we aim at developing an efficient method to estimate the number of communities $K$ when only the adjacency matrix $A$ is known, where $A$ is generated from DCDFM with $K$ communities for arbitrary distribution $\mathcal{F}$ satisfying Equation (\ref{ADCDFM}).
\subsection{Estimation of the number of communities}
Our method for estimating $K$ is closely related to the modularity for signed networks introduced in \cite{gomez2009analysis} and this modularity extends the popular Newman-Girvan modularity matrix \cite{newman2006modularity} from unweighted networks to signed networks. Instead of simply considering signed networks, we extend the modularity developed in \cite{gomez2009analysis} to weighted networks with $A$'s elements being any finite real values by considering indicator functions. We let the $n\times n$ symmetric adjacency matrix $A$ be generated from DCDFM for arbitrary distribution $\mathcal{F}$ satisfying Equation (\ref{ADCDFM}), so we have $A\in\mathbb{R}^{n\times n}$. Let $A^{+}, A^{-}\in\mathbb{R}_{\geq 0}^{n\times n}$ such that $A_{ij}=A^{+}_{ij}-A^{-}_{ij}$, where $A^{+}_{ij}=\mathrm{max}(0,A_{ij})$ and $A^{-}_{ij}=\mathrm{max}(0,-A_{ij})$ for any $i,j\in[n]$. Let $d^{+}$ be the positive degree vector with $i$-th entry $d^{+}_{i}=\sum_{j=1}^{n}A^{+}_{ij}$ and  $d^{-}$ be the negative vector with $i$-th entry $d^{-}_{i}=\sum_{j=1}^{n}A^{-}_{ij}$ for $i\in[n]$. Let $m^{+}=\sum_{i=1}^{n}d^{+}_{i}/2$ and $m^{-}=\sum_{i=1}^{n}d^{-}_{i}/2$. Let $\hat{\ell}$ be a $n\times 1$ node label vector returned by running a community detection method $\mathcal{M}$ on $A$ with $k$ communities such that $\hat{\ell}_{i}$ takes value from $\{1,2,\ldots,k\}$. Based on the community partition $\hat{\ell}$ obtained from the method $\mathcal{M}$, the positive modularity $Q^{+}$ and the negative modularity $Q^{-}$ are defined as
\begin{align*}
Q^{+}=\frac{1}{2m^{+}}\sum_{i=1}^{n}\sum_{j=1}^{n}(A^{+}_{ij}-\frac{d^{+}_{i}d^{+}_{j}}{2m^{+}})\delta(\hat{\ell}_{i},\hat{\ell}_{j})1_{m^{+}>0}, Q^{-}=\frac{1}{2m^{-}}\sum_{i=1}^{n}\sum_{j=1}^{n}(A^{-}_{ij}-\frac{d^{-}_{i}d^{-}_{j}}{2m^{-}})\delta(\hat{\ell}_{i},\hat{\ell}_{j})1_{m^{-}>0},
\end{align*}
where $\delta(\hat{\ell}_{i},\hat{\ell}_{j})$ is the Kronecker delta function, $1_{m^{+}>0}$ and $1_{m^{-}>0}$ are indicator functions such that
\begin{align*}
\delta(\hat{\ell}_{i},\hat{\ell}_{i})=\begin{cases}
1& \mbox{when~} \hat{\ell}_{i}=\hat{\ell}_{i},\\
0, & \mbox{otherwise},
\end{cases},
1_{m^{+}>0}=\begin{cases}
1& \mbox{when~} m^{+}>0,\\
0, & \mbox{otherwise},
\end{cases},
1_{m^{-}>0}=\begin{cases}
1& \mbox{when~} m^{-}>0,\\
0, & \mbox{otherwise},
\end{cases},
\end{align*}
The weighted modularity considered in this article is defined as
\begin{align}\label{Modularity}
Q_{\mathcal{M}}(k)=\frac{2m^{+}}{2m^{+}+2m^{-}}Q^{+}-\frac{2m^{-}}{2m^{+}+2m^{-}}Q^{-}.
\end{align}

When all edge weights are nonnegative such that $m^{-}=0$, the weighted modularity reduces to the Newman-Girvan modularity. When $A$ has both positive and negative entries, the weighted modularity reduces to the modularity introduced in \cite{gomez2009analysis}. The weighted modularity obtained via Equation (\ref{Modularity}) measures the quality of community partition for a weighted network whose adjacency matrix has any finite real elements, and it is more general than the modularity introduced in \cite{gomez2009analysis}. Similar to the Newman-Girvan modularity, a larger weighted modularity $Q_{\mathcal{M}}(k)$ indicates a better community partition.

In Equation (\ref{Modularity}), we write the weighted modularity as a function of the number of communities $k$ and the community detection method $\mathcal{M}$ to emphasize that the weighted modularity may be different for different $k$ or different community detection methods.  We estimate the number of communities $K$ by increasing $k$ until the weighted modularity function in Equation (\ref{Modularity}) does not increase. Recall that Equation (\ref{Modularity}) depends on a community detection method $\mathcal{M}$ and the number of communities $k$. Suppose there is a cardinality choice of $K$ such that $K$ locates in $\{1,2,\ldots, K_{0}\}$.  For a community detection algorithm $\mathcal{M}$, our strategy for estimating $K$ is
\begin{align}\label{SpectralModularity}
\hat{K}_{\mathcal{M}}=\underset{k\in[K_{0}]}{\mathrm{arg~max}}Q_{\mathcal{M}}(k).
\end{align}

In this paper, to estimate the number of communities for weighted networks generated from DCDFM, we choose the method $\mathcal{M}$ as the nDFA algorithm designed in \cite{DCDFM} because nDFA enjoys consistent estimation of community memberships under the DCDFM model and it is computationally fast. For convenience, when $\mathcal{M}$ is the nDFA algorithm, we call our method for estimating $K$ via Equation (\ref{SpectralModularity}) as nDFAwm, where ``wm'' means weighted modularity. The details of the nDFA algorithm \cite{DCDFM} are written below.

Input: $A, k$. Output: $\hat{\ell}$.
\begin{itemize}
  \item Let $\tilde{A}=\hat{U}\hat{\Lambda}\hat{U}'$ be the top-$k$ eigendecomposition of $A$.
  \item Let the $n\times k$ matrix $\hat{U}_{*}$ be the row normalization of $\hat{U}$ such that $\hat{U}_{*}(i,:)=\frac{\hat{U}(i,:)}{\|\hat{U}(i,:)\|_{F}}$ for $i\in[n]$.
  \item Apply k-means algorithm on all rows of $\hat{U}_{*}$ with k clusters to obtain $\hat{\ell}$.
\end{itemize}
\section{Simulations}\label{Simulations}
In this section, we compare our nDFAwm with three model-free methods in the literature for estimating the number of communities: the modularity eigengap (ME for short) method proposed in \cite{budel2020detecting}, the non-backtracking (NB) method designed in \cite{le2022estimating}, and the Bethe Hessian matrix-based method BHac developed in \cite{le2022estimating}. For each parameter setting considered in this section, we report the Accuracy rate over 100 repetitions for each method, where the Accuracy rate is the fraction of times that the estimated number of clusters $\hat{K}$ equals the true number of clusters $K$.

To generate simulated weighted networks from DCDFM, first, we need to define $n, K, \theta, Z$, and $P$. For $n$, unless specified, we let $n=50K$. For $Z$, we let each node belong to one of the $K$ clusters with equal probability, i.e., there are around 50 nodes in each cluster. For $\theta$, unless specified, we let $\theta_{i}=\mathrm{rand}(1)\sqrt{\rho}$, where the positive number $\rho$ controls network sparsity and $\mathrm{rand}(1)$ is a random number drawn from the uniform distribution in the interval $(0,1)$. We set $n, K, P$, and $\rho$ independently for each simulation. After setting these model parameters, we generate $A$ under DCDFM for several distributions $\mathcal{F}$ satisfying Equation (\ref{ADCDFM}). For our nDFAwm, we set $K_{c}=20$ since the largest $K$ in our simulations is 6. In this paper, we consider Bernoulli, Binomial, Poisson, Geometrical, Exponential, Normal, Laplace, and Uniform distributions, where details on probability mass function or probability density function of these distributions can be found in \url{http://www.stat.rice.edu/~dobelman/courses/texts/distributions.c&b.pdf}. Meanwhile, we also consider the signed network case in our simulation studies.
\subsection{Bernoulli distribution}
When $\mathcal{F}$ is Bernoulli distribution such that $A_{ij}\sim \mathrm{Bernoulli}(\Omega_{ij})$, i.e., $A_{ij}\in\{0,1\}$ for $i,j\in[n]$ and DCDFM reduces to DCSBM for this case. By the property of Bernoulli distribution, $\mathbb{E}[A_{ij}]=\Omega_{ij}$ satisfies Equation (\ref{ADCDFM}) and $\Omega_{ij}$ is a probability ranging in $[0,1]$. So, $\rho$'s range is $(0,1]$, and all elements of $P$ should be nonnegative. For Bernoulli distribution, we consider the following simulations.
\begin{figure}
\centering
\subfigure[Experiment 1 (a)]{\includegraphics[width=0.24\textwidth]{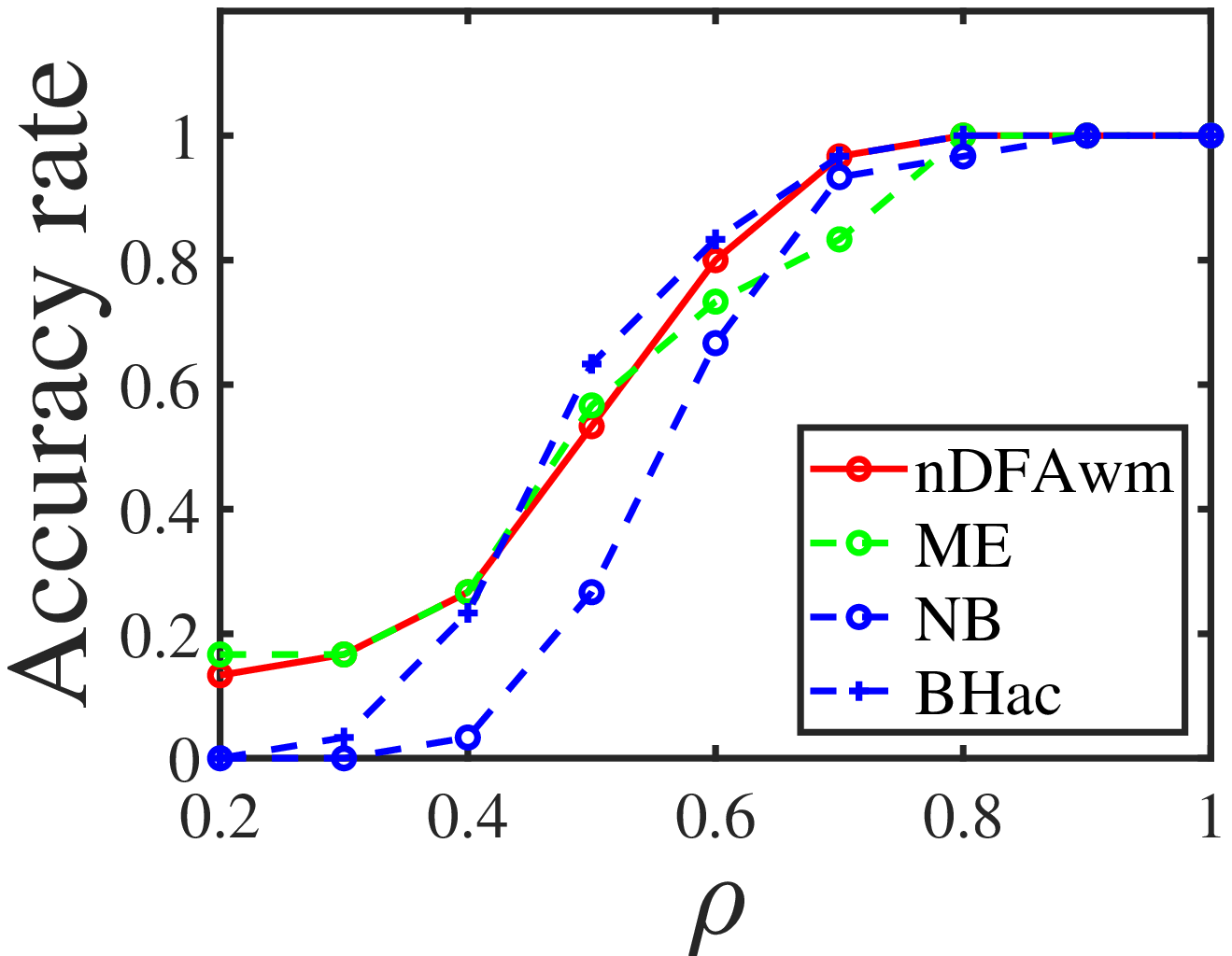}}
\subfigure[Experiment 1 (b)]{\includegraphics[width=0.24\textwidth]{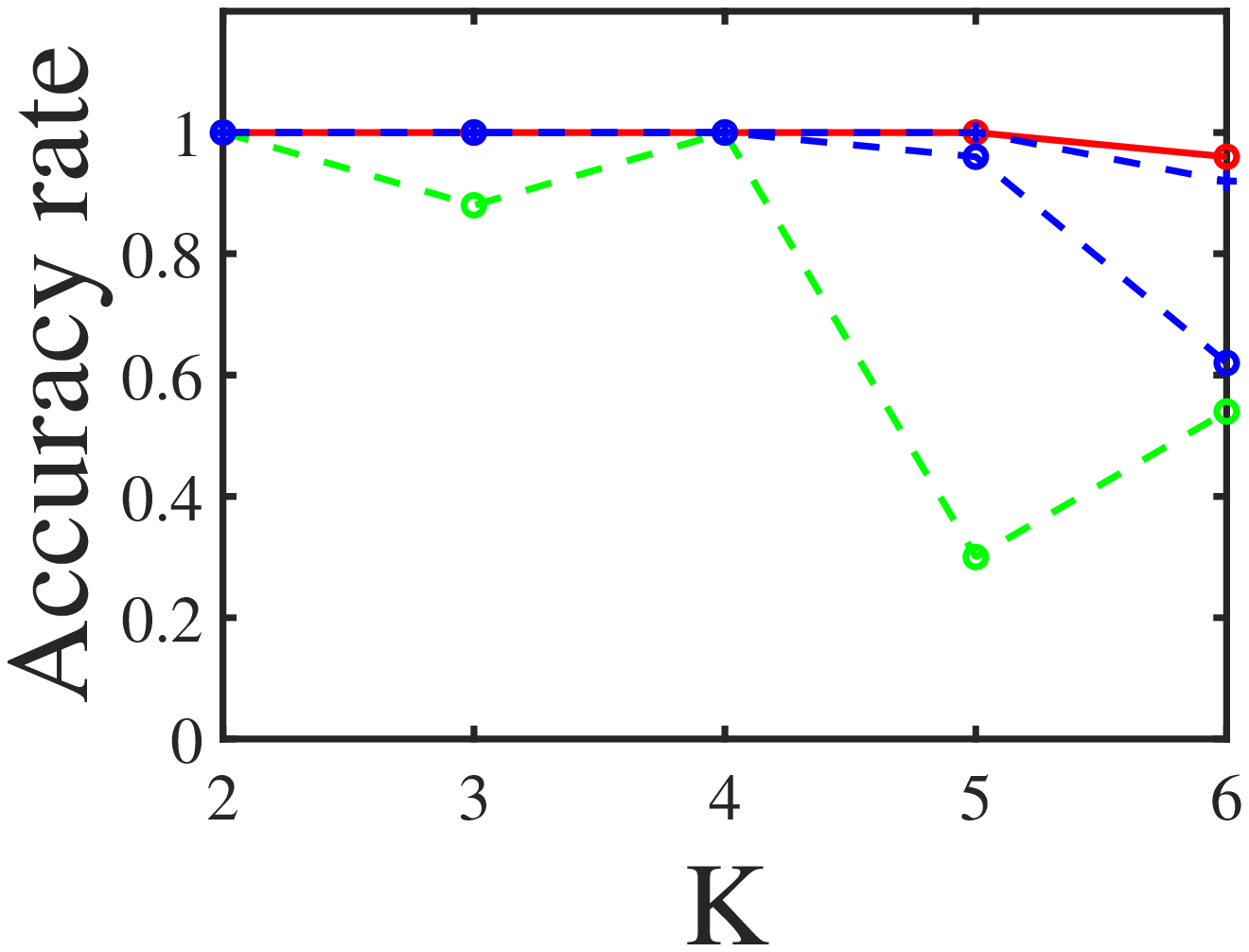}}
\subfigure[Experiment 1 (c)]{\includegraphics[width=0.24\textwidth]{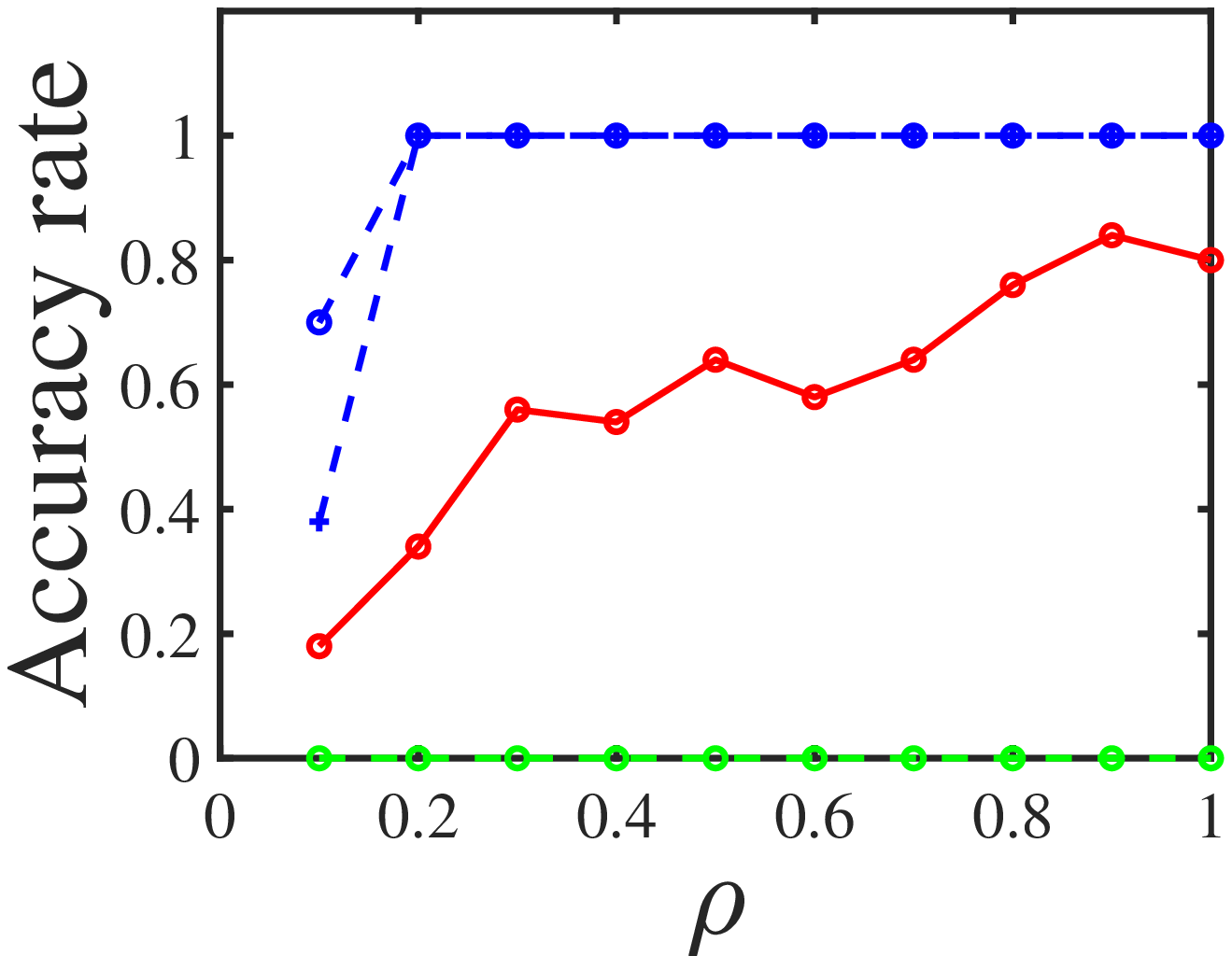}}
\subfigure[Experiment 1 (d)]{\includegraphics[width=0.24\textwidth]{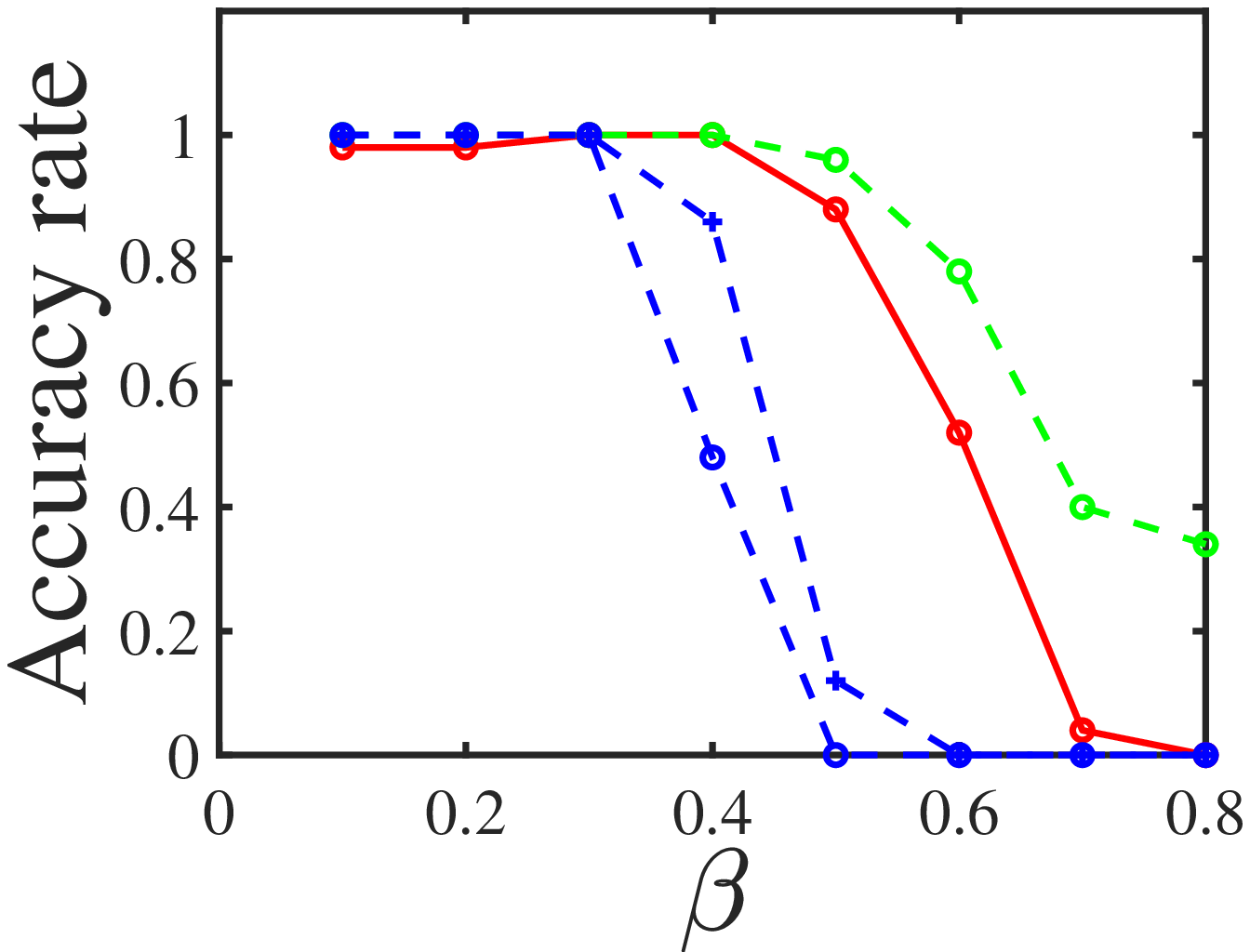}}
\caption{Bernoulli distribution.}
\label{S1} 
\end{figure}

\texttt{Experiment 1 (a): changing $\rho$.} Let $K=3$ and $P$ be
\[P=\begin{bmatrix}
    1&0.2&0.3\\
    0.2&0.8&0.2\\
    0.3&0.2&0.9\\
\end{bmatrix}.\]
Let $\rho$ range in $\{0.2, 0.3, \ldots, 1\}$.

\texttt{Experiment 1 (b): changing $K$.} Let $P$'s diagonal entries be 1 and off-diagonal entries be 0.2. Let $\rho=0.9$ and $K$ range in $\{2,3,\ldots, 6\}$.

\texttt{Experiment 1 (c): changing $\rho$ when $K=1$.} Let $K=1, P=1$, and $\rho$ range in $\{0.1, 0.2, \ldots, 1\}$.

\texttt{Experiment 1 (d): connectivity across communities.} Let $K=2, \rho=1$, $P$'s diagonal entries be 1, $P$'s off-diagonal entries be $\beta$, and $\beta$ range in $\{0.1, 0.2, \ldots, 0.8\}$.

Figure \ref{S1} shows the Accuracy rate of Experiment 1. Panel (a) of Figure \ref{S1} says that as the network becomes denser, all methods provide more accurate estimations of the number of clusters. For Experiment 1 (a), all methods perform similarly. For Experiment 1 (b), from panel (b) of Figure \ref{S1}, we see that our nDFAwm performs best. From panel (c) of Figure \ref{S1}, we see that our nDFAwm performs poorer than NB and BHac while ME fails to work. Meanwhile, except ME, all methods perform better as the network becomes denser for Experiment 1 (c). From panel (d) of Figure \ref{S1}, we see that all methods perform poorer as the off-diagonal entries of $P$ are closer to the diagonal entries and our nDFAwm performs slightly poorer than ME while it outperforms NB and BHac.
\subsection{Binomial distribution}
When $\mathcal{F}$ is Binomial distribution such that $A_{ij}\sim \mathrm{Binomial}(m,\frac{\Omega_{ij}}{m})$ for any positive integer $m$, i.e., $A_{ij}\in\{0,1,2,\ldots,m\}$ for $i,j\in[n]$. By the property of Binomial distribution, $\mathbb{E}[A_{ij}]=\Omega_{ij}$ satisfies Equation (\ref{ADCDFM}) and $\frac{\Omega_{ij}}{m}$ is a probability ranging in $[0,1]$. So, $\rho$'s range is $(0,m]$ and all elements of $P$ should be nonnegative.
\begin{figure}
\centering
\subfigure[Experiment 2 (a)]{\includegraphics[width=0.24\textwidth]{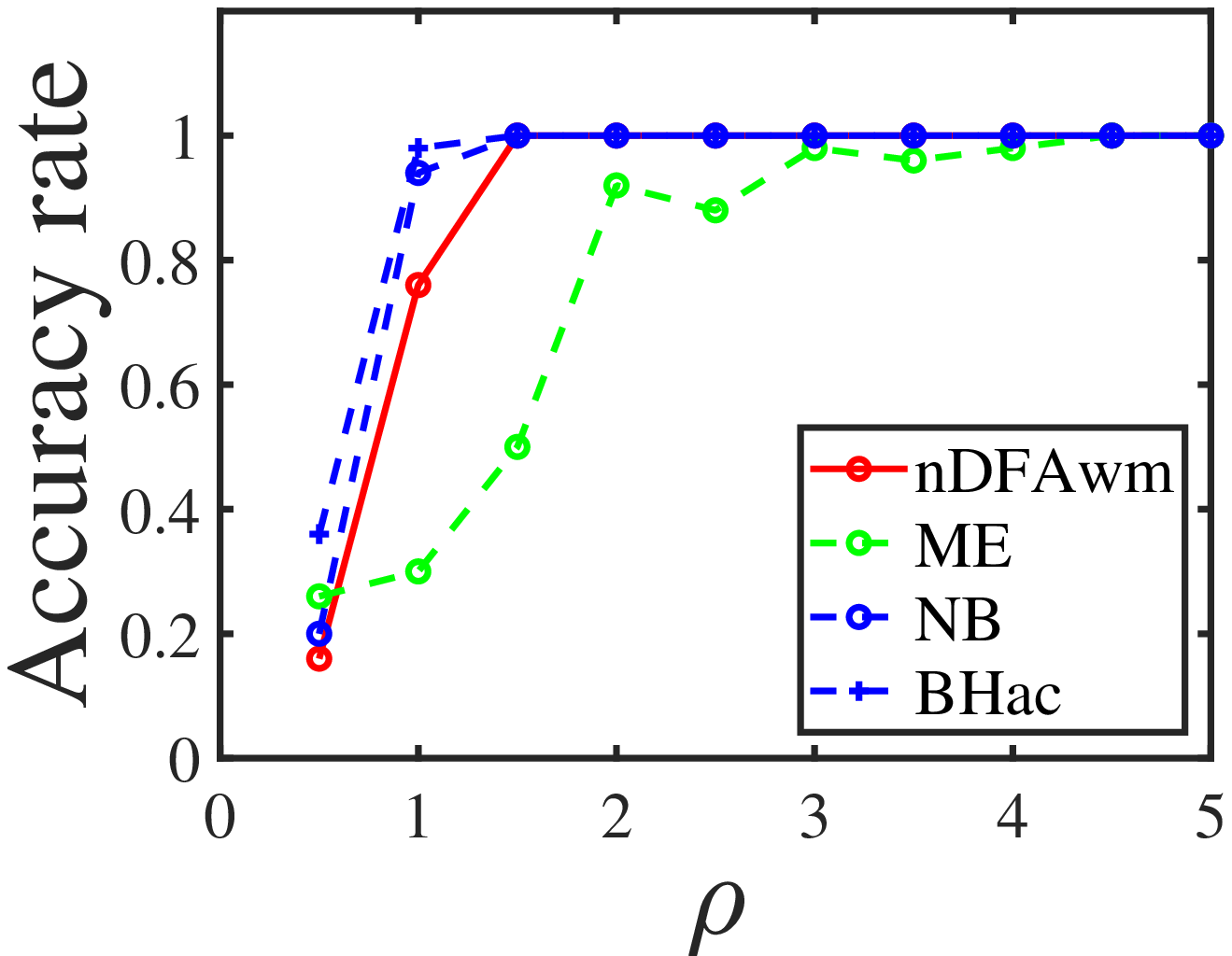}}
\subfigure[Experiment 2 (b)]{\includegraphics[width=0.24\textwidth]{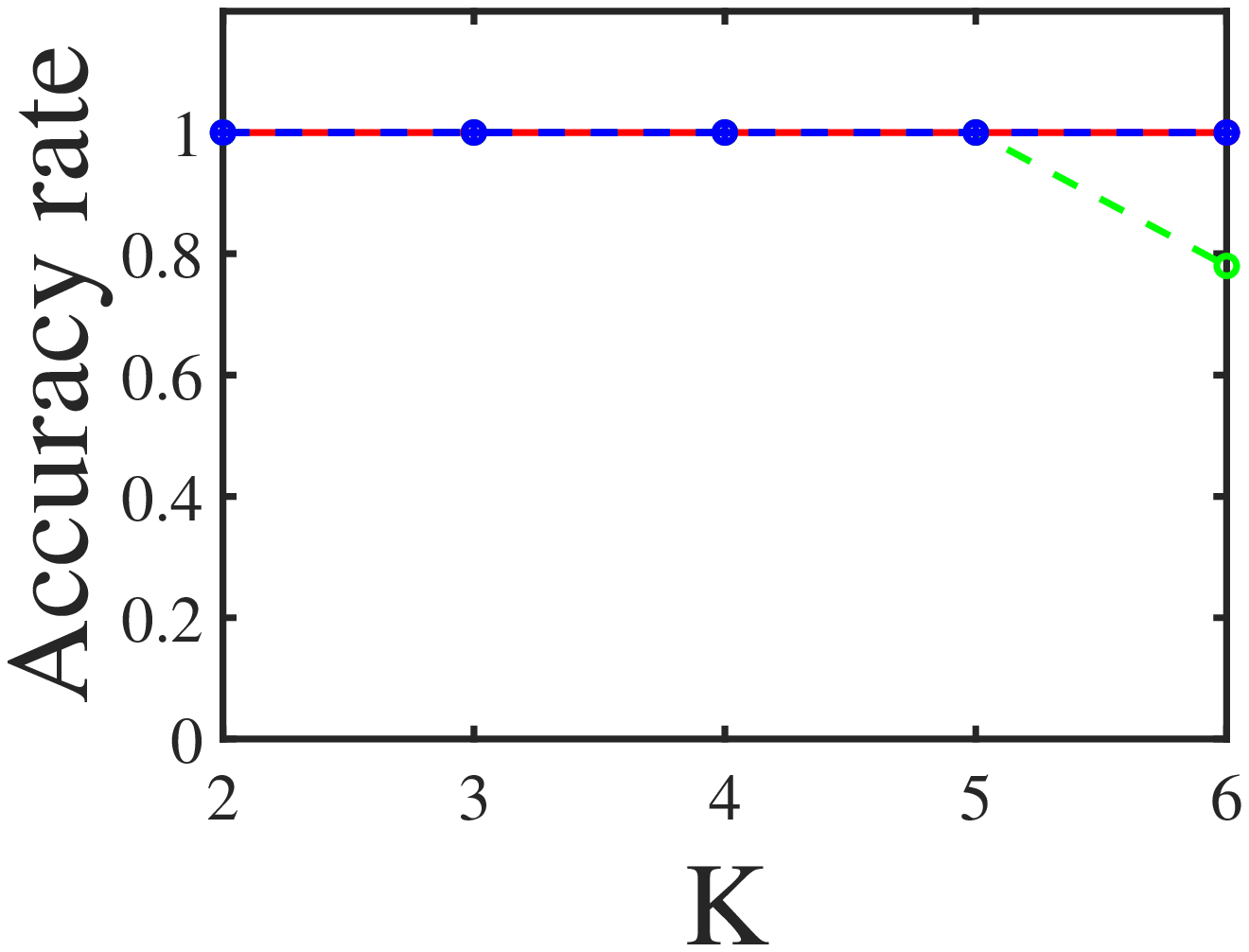}}
\subfigure[Experiment 2 (c)]{\includegraphics[width=0.24\textwidth]{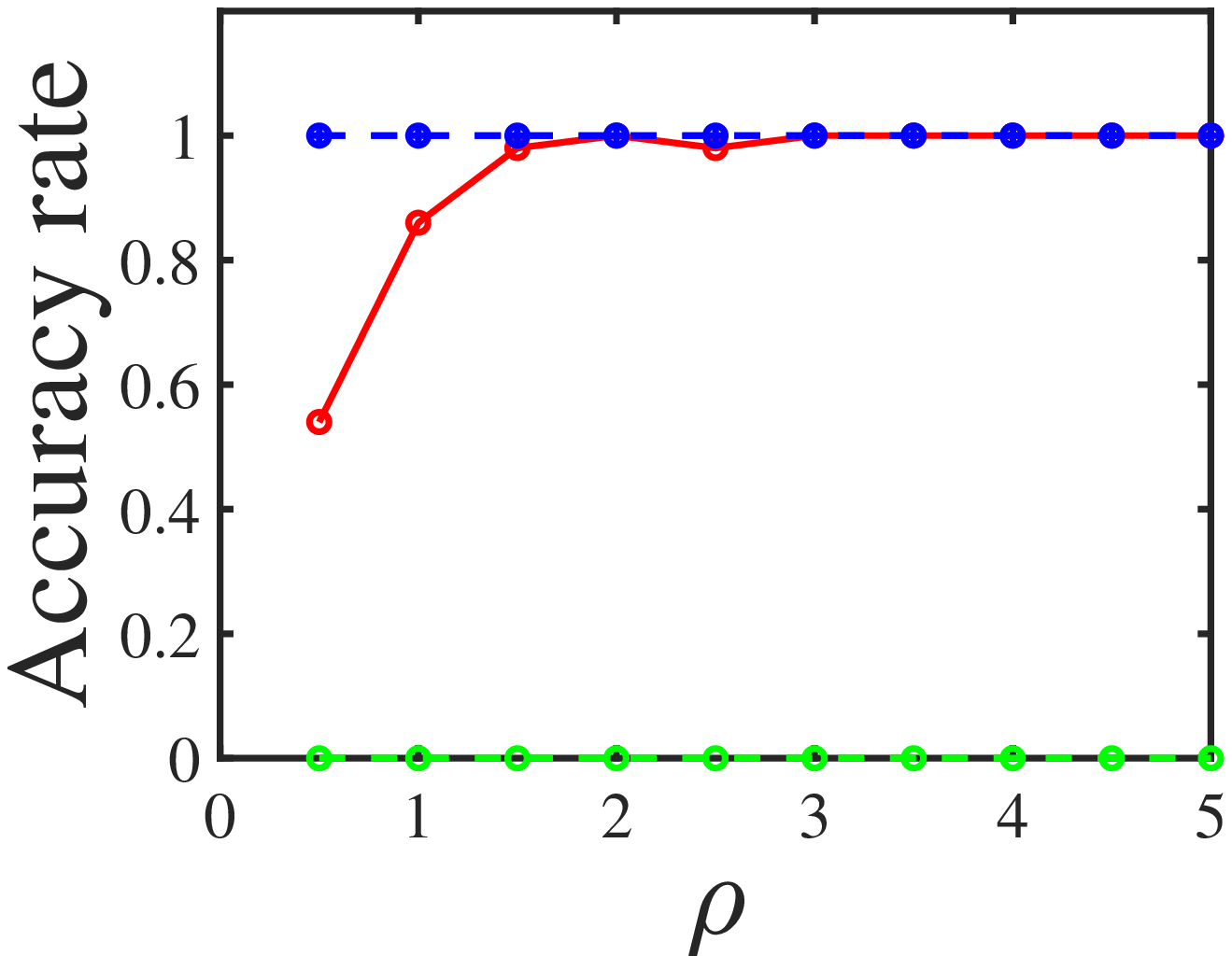}}
\subfigure[Experiment 2 (d)]{\includegraphics[width=0.24\textwidth]{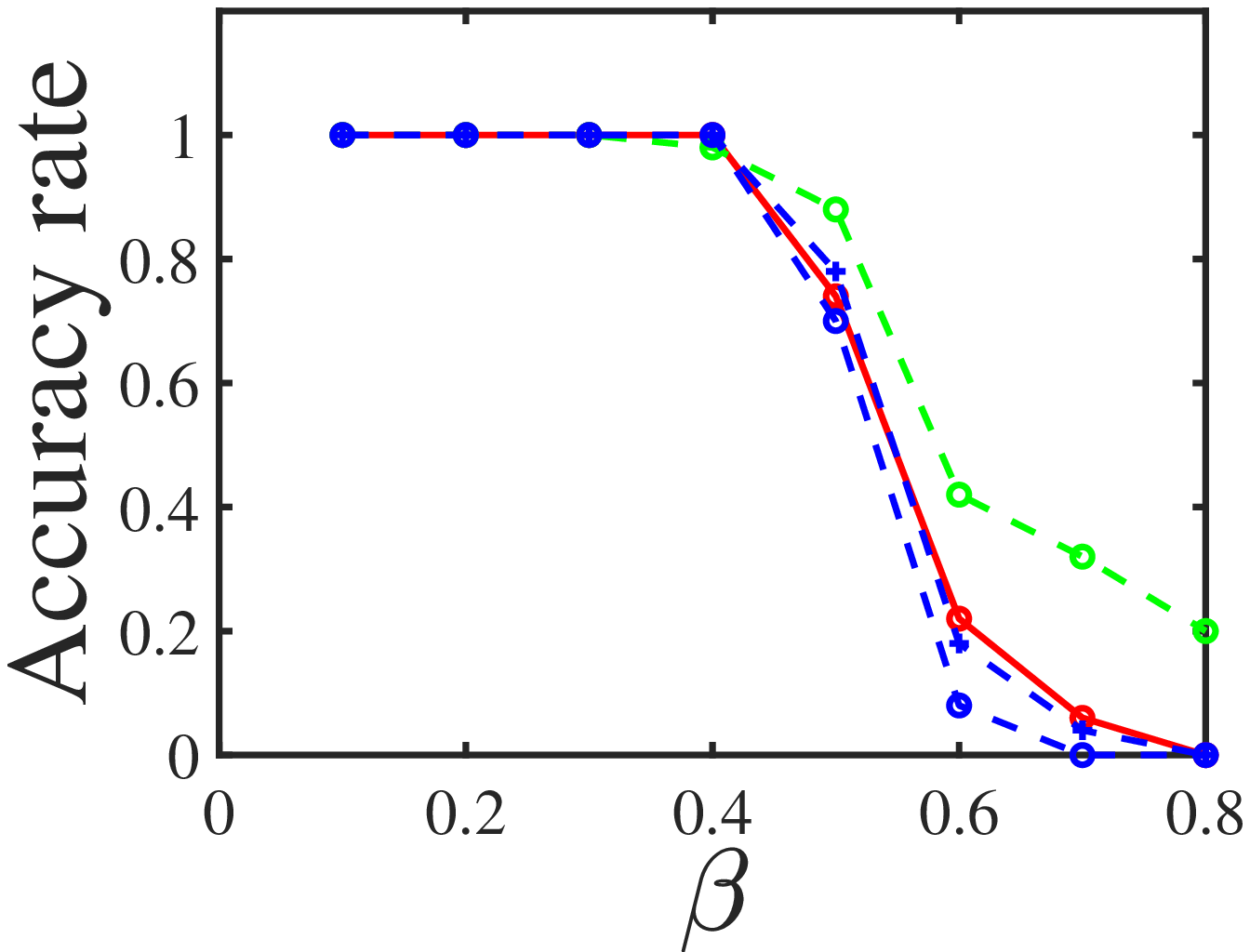}}
\caption{Binomial distribution.}
\label{S2} 
\end{figure}

\texttt{Experiment 2 (a): changing $\rho$.} Let $K=3, m=5$, and $P$ be the same as that of Experiment 1 (a). Let $\rho$ range in $\{0.5, 1, \ldots, 5\}$.

\texttt{Experiment 2 (b): changing $K$.} Let $P$ be the same as Experiment 1 (b), $\rho=2, m=5$, and $K$ range in $\{2,3,\ldots, 6\}$.

\texttt{Experiment 2 (c): changing $\rho$ when $K=1$.} Let $K=1, P=1, m=5$, and $\rho$ range in $\{0.5, 1, \ldots, 5\}$.

\texttt{Experiment 2 (d): connectivity across communities.} Let $K=2, \rho=1, m=5$, and $P$ be the same as Experiment 1 (d).

Figure \ref{S2} shows the Accuracy rate of Experiment 2. For Experiments 2 (a), 2 (b), and 2 (c), the results are similar to that of Experiments 1 (a), 1 (b), and 1 (c), respectively, and we omit the analysis here. For Experiment 2 (d), panel (d) of Figure \ref{S2} says that our nDFAwm perform similarly to NB and BHac while ME performs best.
\subsection{Poisson distribution}
When $\mathcal{F}$ is Poisson distribution such that $A_{ij}\sim \mathrm{Poisson}(\Omega_{ij})$, i.e., $A_{ij}$ is a nonnegative integer for $i,j\in[n]$. By the property of Poisson distribution, $\mathbb{E}[A_{ij}]=\Omega_{ij}$ satisfies Equation (\ref{ADCDFM}) and $\Omega_{ij}$ is nonnegative. So, $\rho$'s range is $(0,+\infty)$ and all elements of $P$ should be nonnegative.
\begin{figure}
\centering
\subfigure[Experiment 3 (a)]{\includegraphics[width=0.24\textwidth]{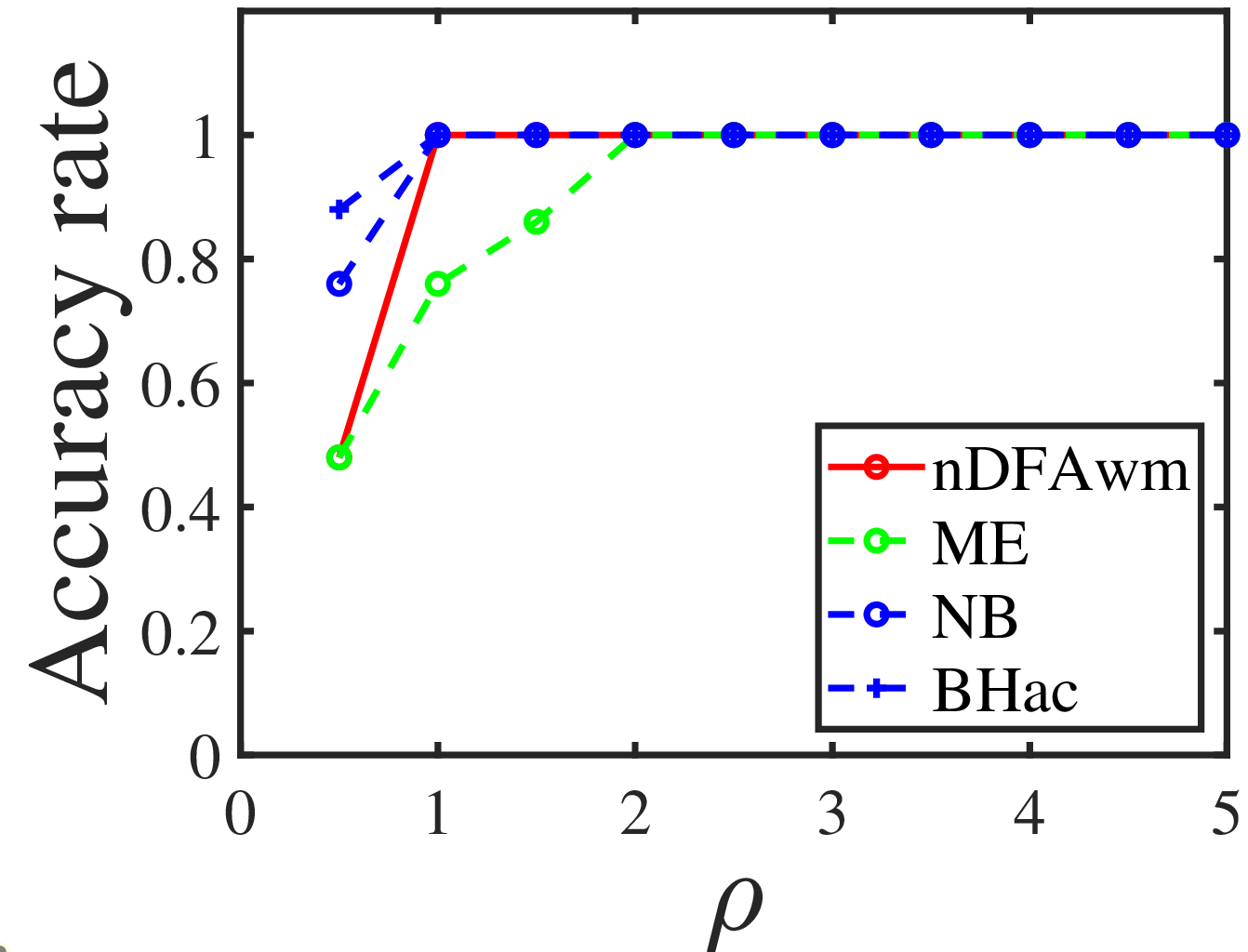}}
\subfigure[Experiment 3 (b)]{\includegraphics[width=0.24\textwidth]{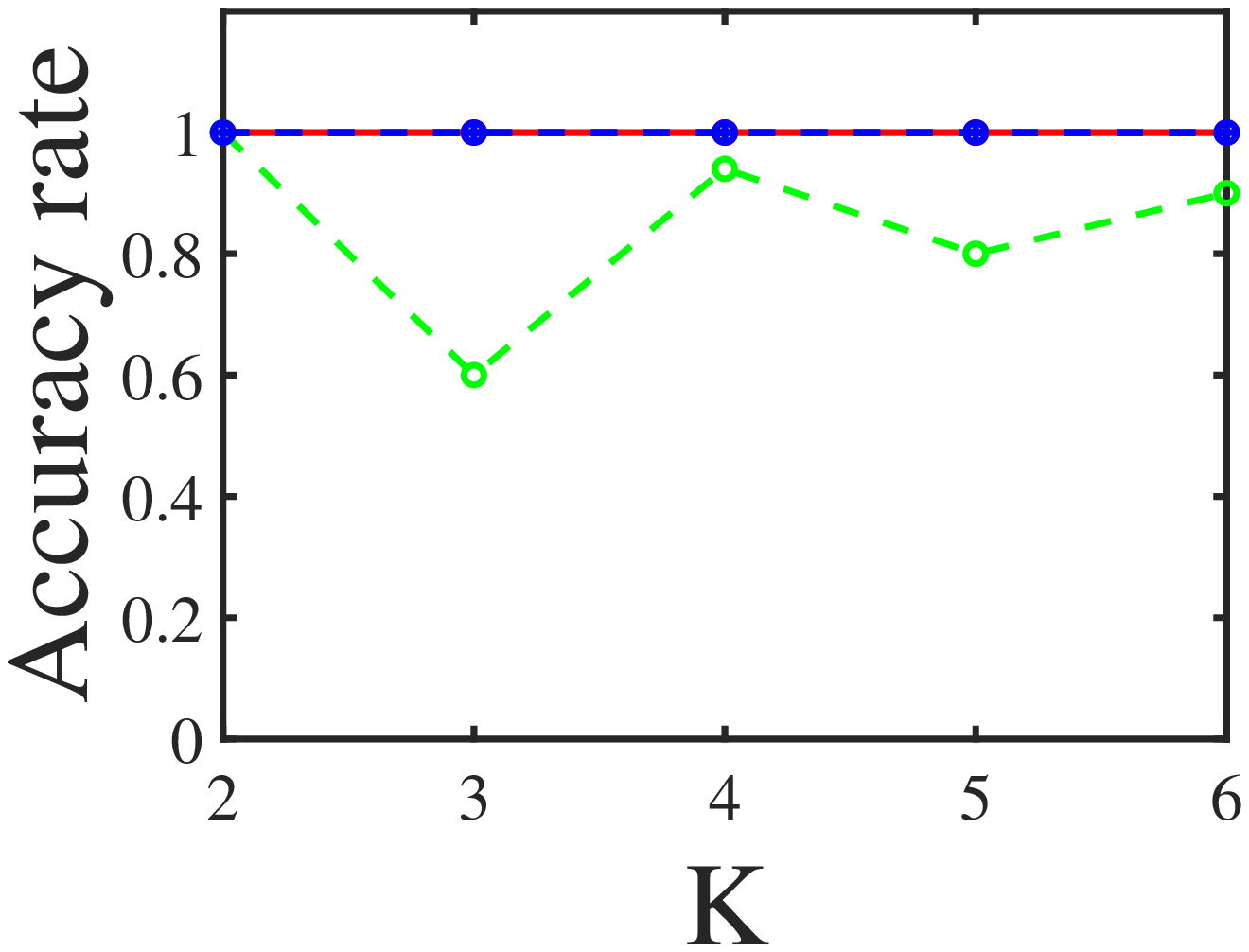}}
\subfigure[Experiment 3 (c)]{\includegraphics[width=0.24\textwidth]{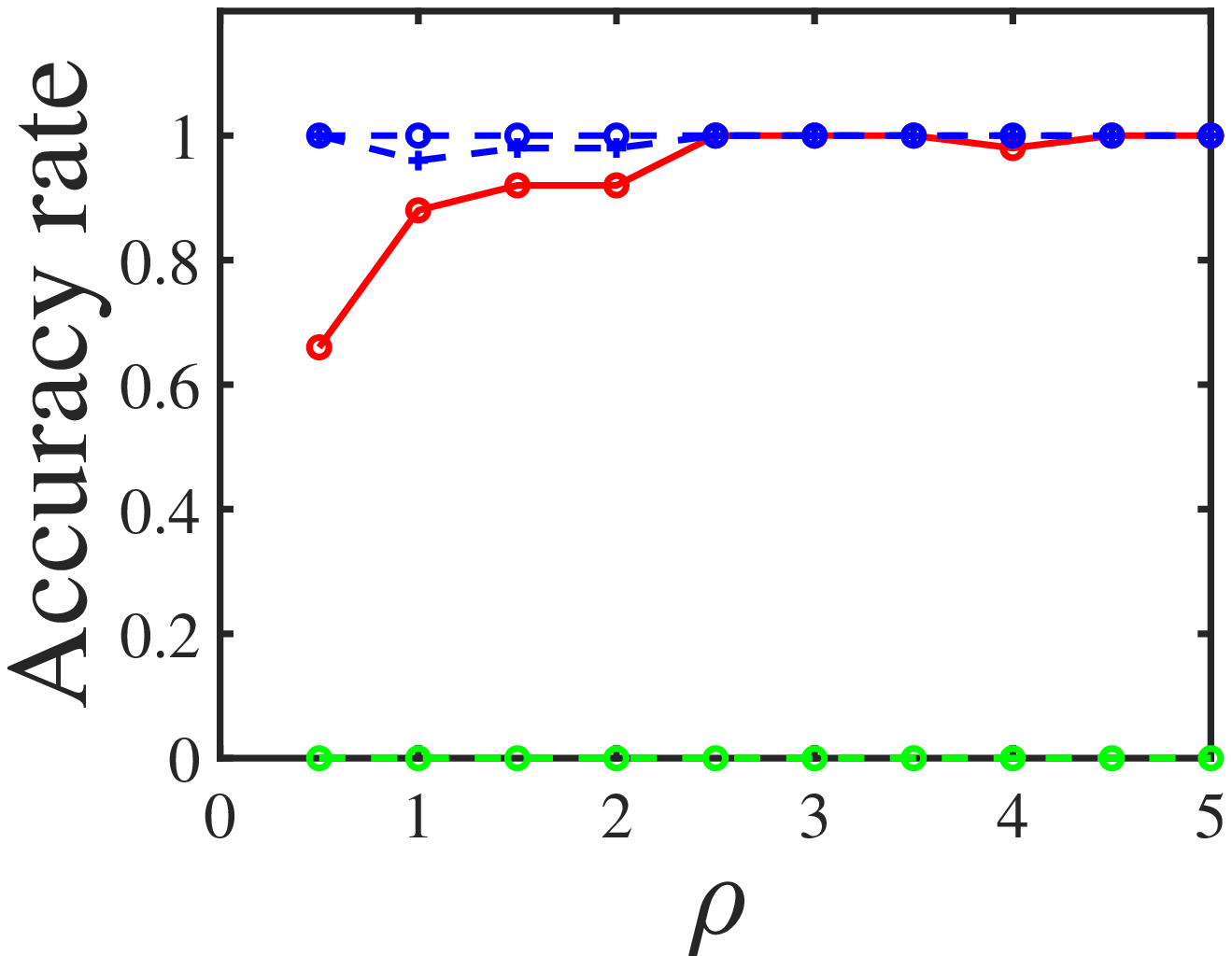}}
\subfigure[Experiment 3 (d)]{\includegraphics[width=0.24\textwidth]{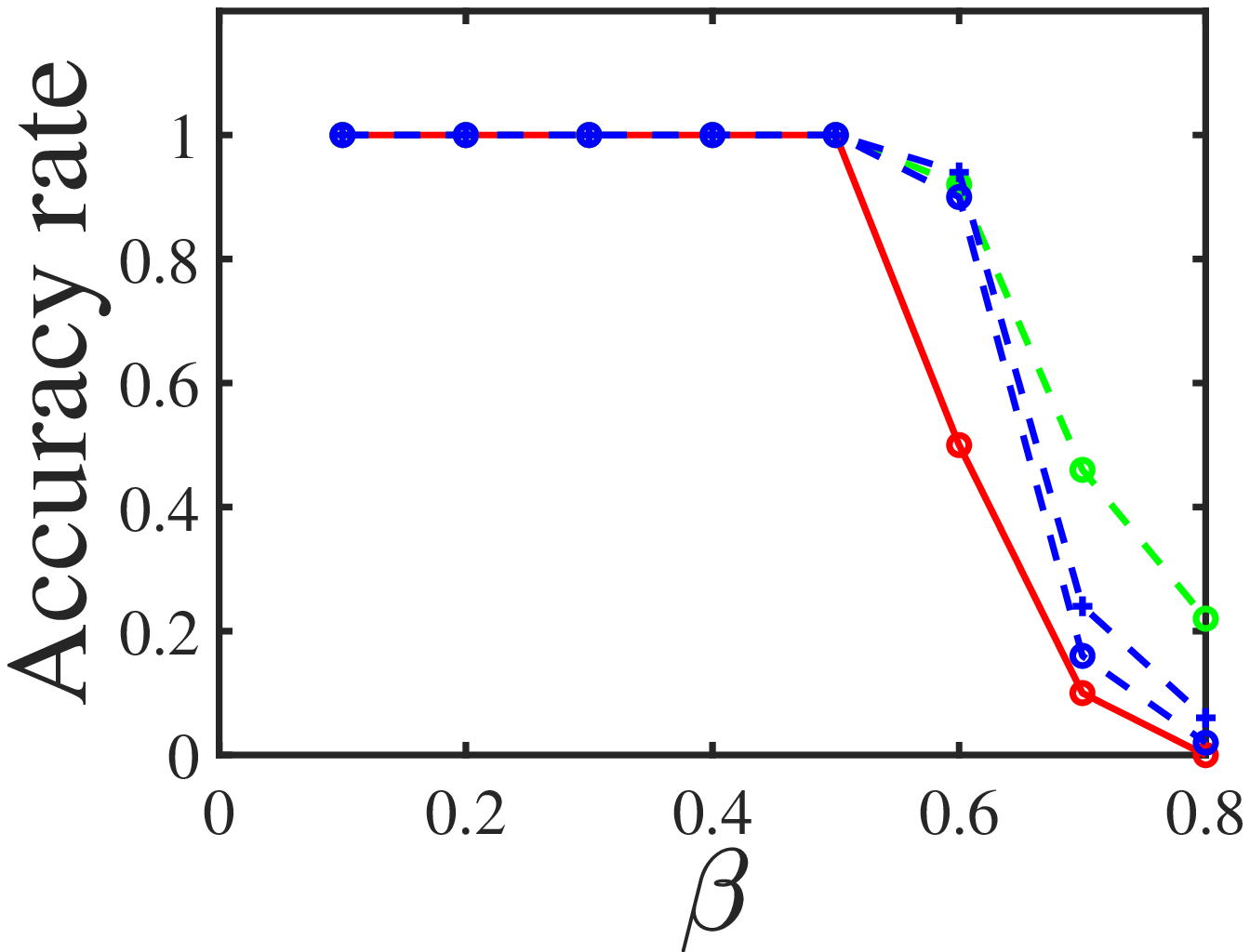}}
\caption{Poisson distribution.}
\label{S3} 
\end{figure}

\texttt{Experiment 3 (a): changing $\rho$.} Let $K=3$ and $P$ be the same as that of Experiment 1 (a). Let $\rho$ range in $\{0.5, 1, \ldots, 5\}$.

\texttt{Experiment 3 (b): changing $K$.} Let $P$ be the same as Experiment 1 (b), $\rho=2$, and $K$ range in $\{2,3,\ldots, 6\}$.

\texttt{Experiment 3 (c): changing $\rho$ when $K=1$.} Let $K=1, P=1$, and $\rho$ range in $\{0.5, 1, \ldots, 5\}$.

\texttt{Experiment 3 (d): connectivity across communities.} Let $K=2, \rho=2$, and $P$ be the same as Experiment 1 (d).

Figure \ref{S3} shows the Accuracy rate of Experiment 3. The results are similar to that of Experiment 2, and we omit the analysis here.
\subsection{Geometric distribution}
When $\mathcal{F}$ is a Geometric distribution such that $A_{ij}\sim\mathrm{Geometric}(\frac{1}{\Omega_{ij}})$, i.e., $A_{ij}$ is positive integer for $i,j\in[n]$. For Geometric distribution, since $\mathbb{P}(A_{ij}=m)=\frac{1}{\Omega_{ij}}(1-\frac{1}{\Omega_{ij}})^{m-1}$ for $m=1,2,\ldots,$ and $ 0<\frac{1}{\Omega_{ij}}\leq 1$, all elements of $P$ must be positive. By the property of Geometric distribution, we have $\mathbb{E}[A_{ij}]=\Omega_{ij}$ satisfying Equation (\ref{ADCDFM}).  For convenience, we let $\theta_{i}=\sqrt{\rho}$ for $i\in[n]$ to make DCDFM reduce to DFM for this case. Then, we have $\Omega=\rho ZPZ'$. Since $\Omega_{ij}\geq1$ for $i,j\in[n]$, we have $\rho\mathrm{min}_{k,l\in[K]}P_{kl}\geq1$.
\begin{figure}
\centering
\subfigure[Experiment 4 (a)]{\includegraphics[width=0.24\textwidth]{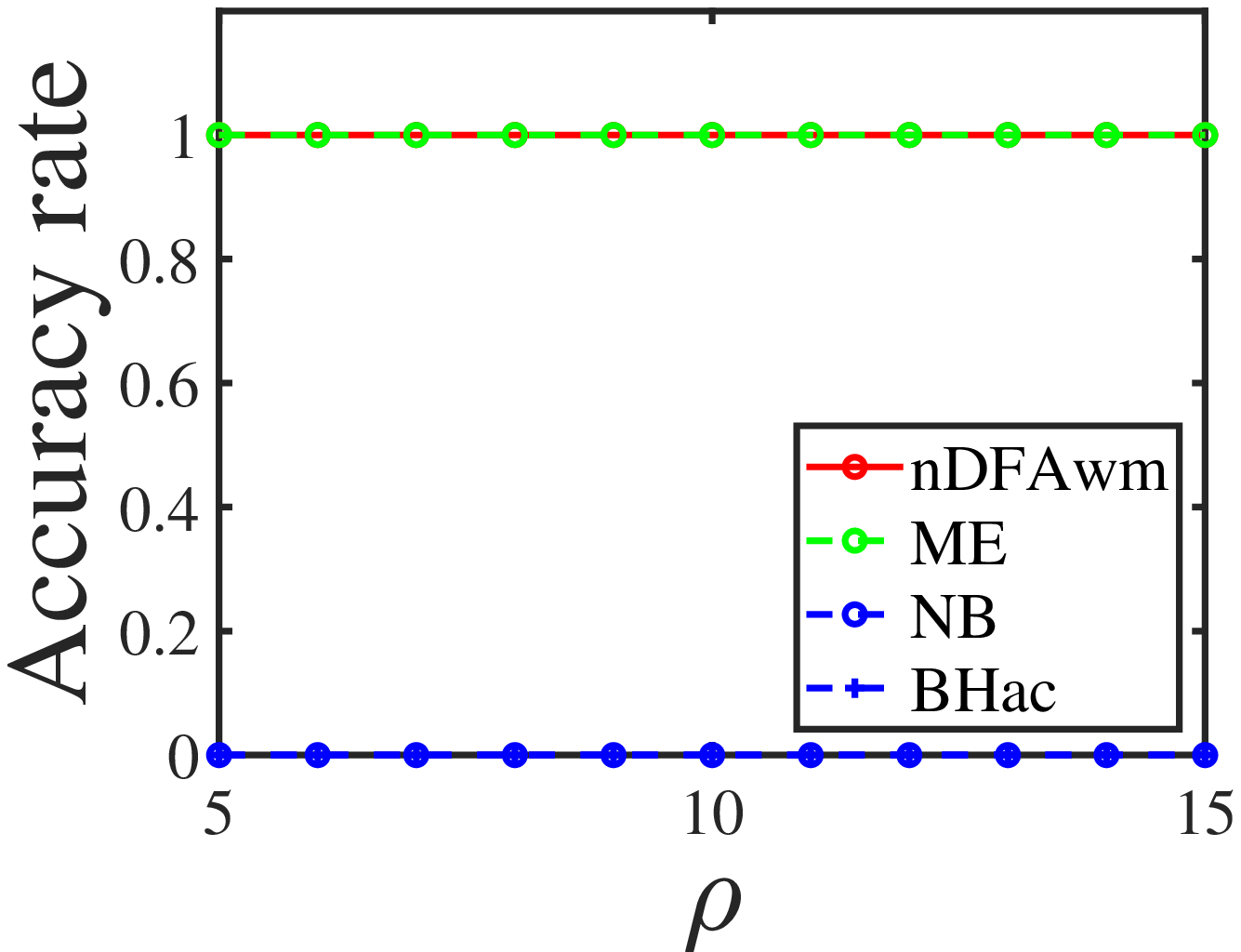}}
\subfigure[Experiment 4 (b)]{\includegraphics[width=0.24\textwidth]{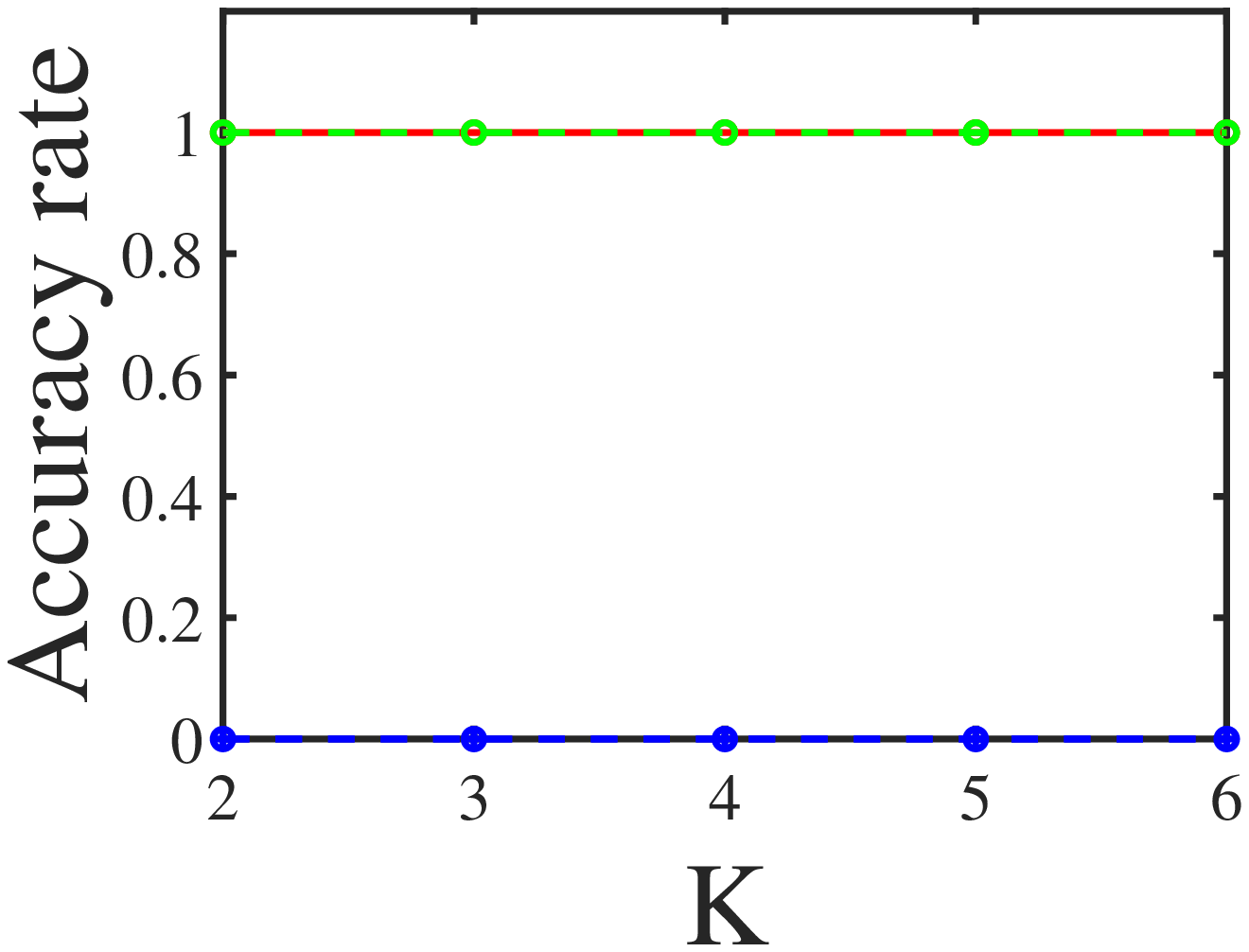}}
\subfigure[Experiment 4 (c)]{\includegraphics[width=0.24\textwidth]{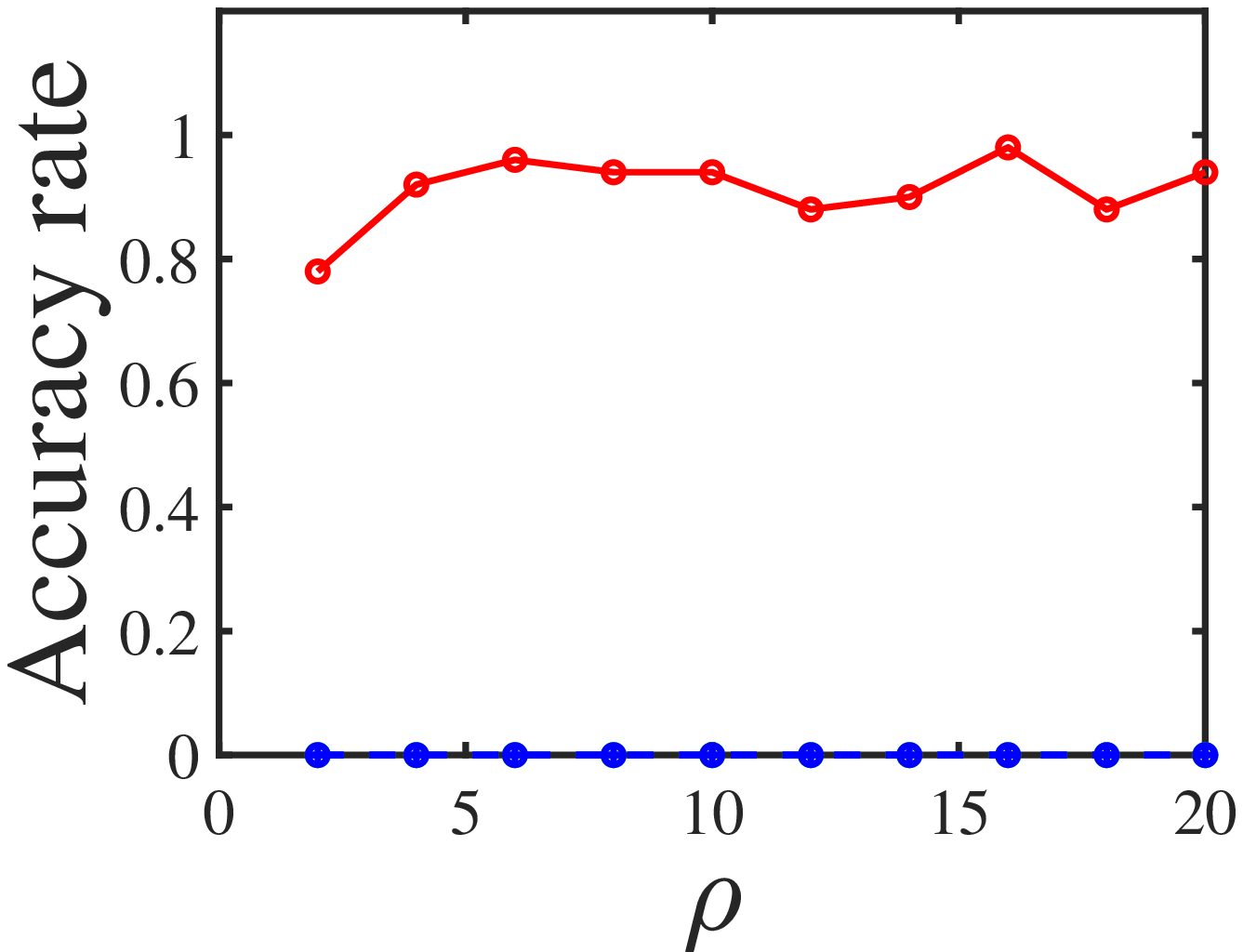}}
\subfigure[Experiment 4 (d)]{\includegraphics[width=0.24\textwidth]{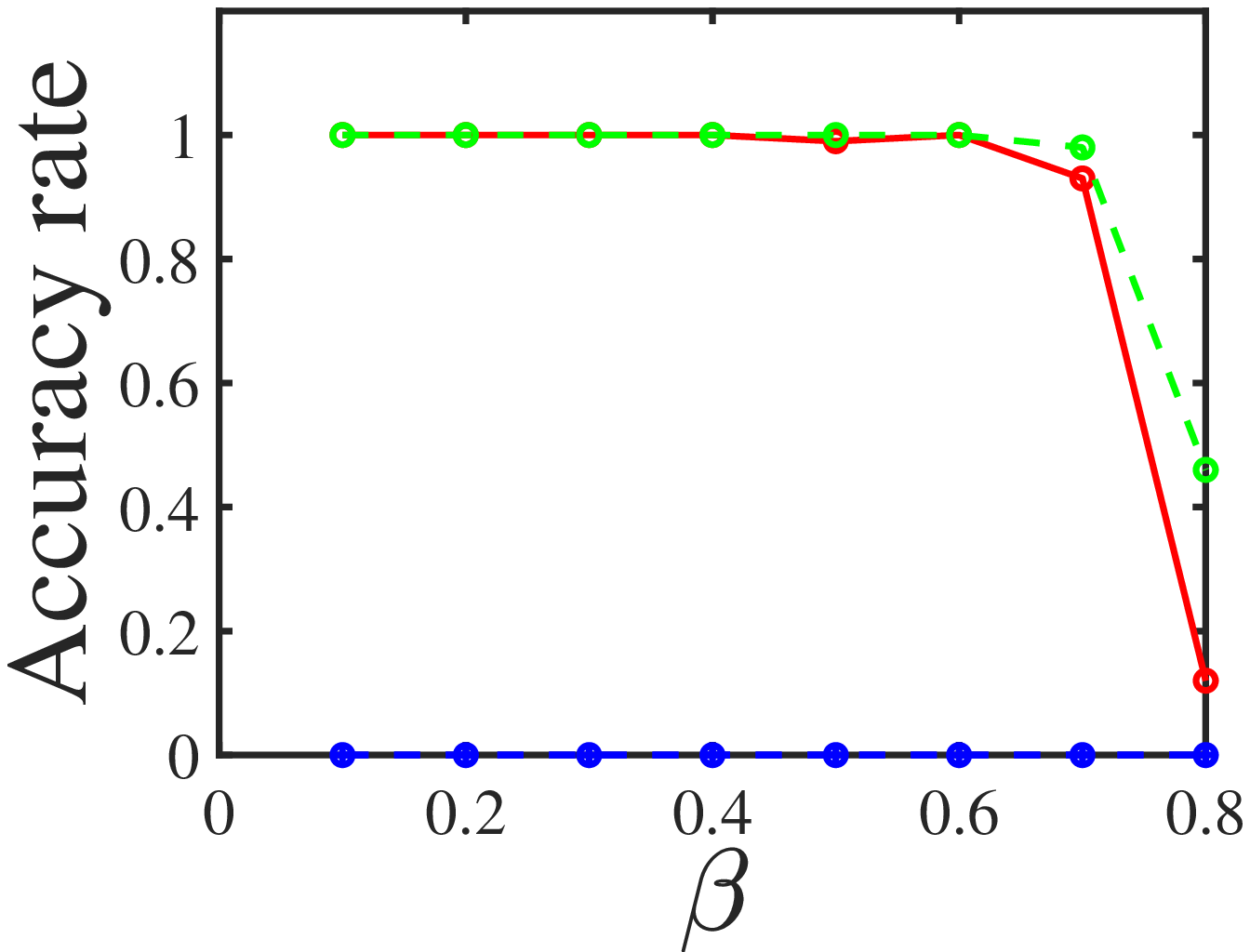}}
\caption{Geometric distribution.}
\label{S4} 
\end{figure}

\texttt{Experiment 4 (a): changing $\rho$.} Let $K=3$ and $P$ be the same as that of Experiment 1 (a). Let $\rho$ range in $\{5,6, \ldots, 15\}$.

\texttt{Experiment 4 (b): changing $K$.} Let $P$ be the same as Experiment 1 (b), $\rho=10$, and $K$ range in $\{2,3,\ldots, 6\}$.

\texttt{Experiment 4 (c): changing $\rho$ when $K=1$.} Let $K=1, P=1$, and $\rho$ range in $\{2, 4, \ldots, 20\}$.

\texttt{Experiment 4 (d): connectivity across communities.} Let $K=2, \rho= 10$, and $P$ be the same as Experiment 1 (d).

Figure \ref{S4} shows the Accuracy rate of Experiment 4. Unlike Experiments 1-3, the numerical results of Experiment 4 say that our nDFAwm successfully estimates the number of communities for all cases while NB and BHac fail to work when the network is generated from Geometric distribution under the DCDFM model. For the method ME, it fails to work when the true $K$ is 1 and it performs similarly to our nDFAwm for other cases.
\subsection{Exponential distribution}
When $\mathcal{F}$ is a Exponential distribution such that $A_{ij}\sim\mathrm{Exponential}(\frac{1}{\Omega_{ij}})$, i.e., $A_{ij}\in\mathbb{R}_{+}$ for $i,j\in[n]$. For Exponential distribution, since $\frac{1}{\Omega_{ij}}>0$, all elements of $P$ must be positive and $\rho$ range in $(0,+\infty)$. By the property of Exponential distribution, $\mathbb{E}[A_{ij}]=\Omega_{ij}$ satisfies Equation (\ref{ADCDFM}).
\begin{figure}
\centering
\subfigure[Experiment 5 (a)]{\includegraphics[width=0.24\textwidth]{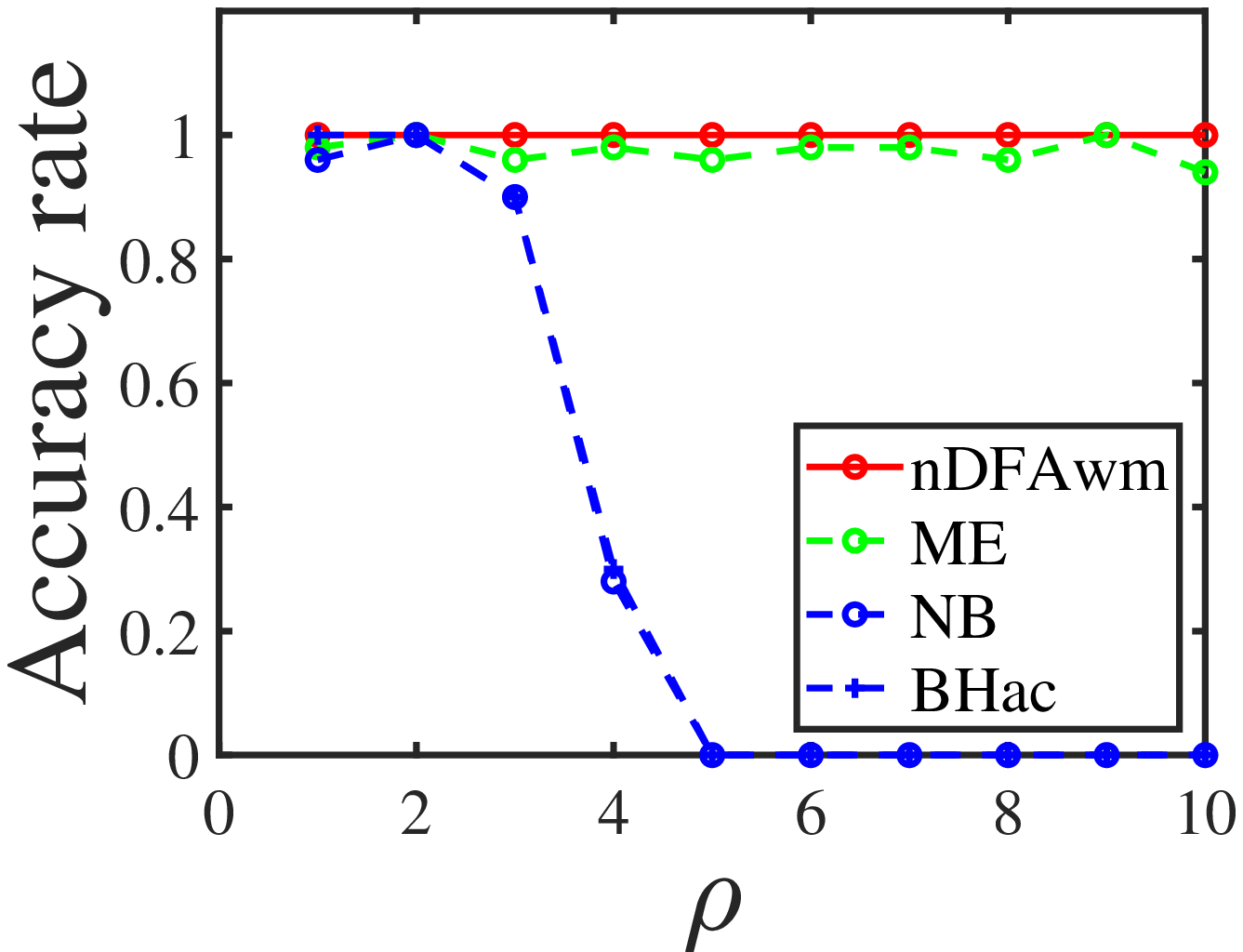}}
\subfigure[Experiment 5 (b)]{\includegraphics[width=0.24\textwidth]{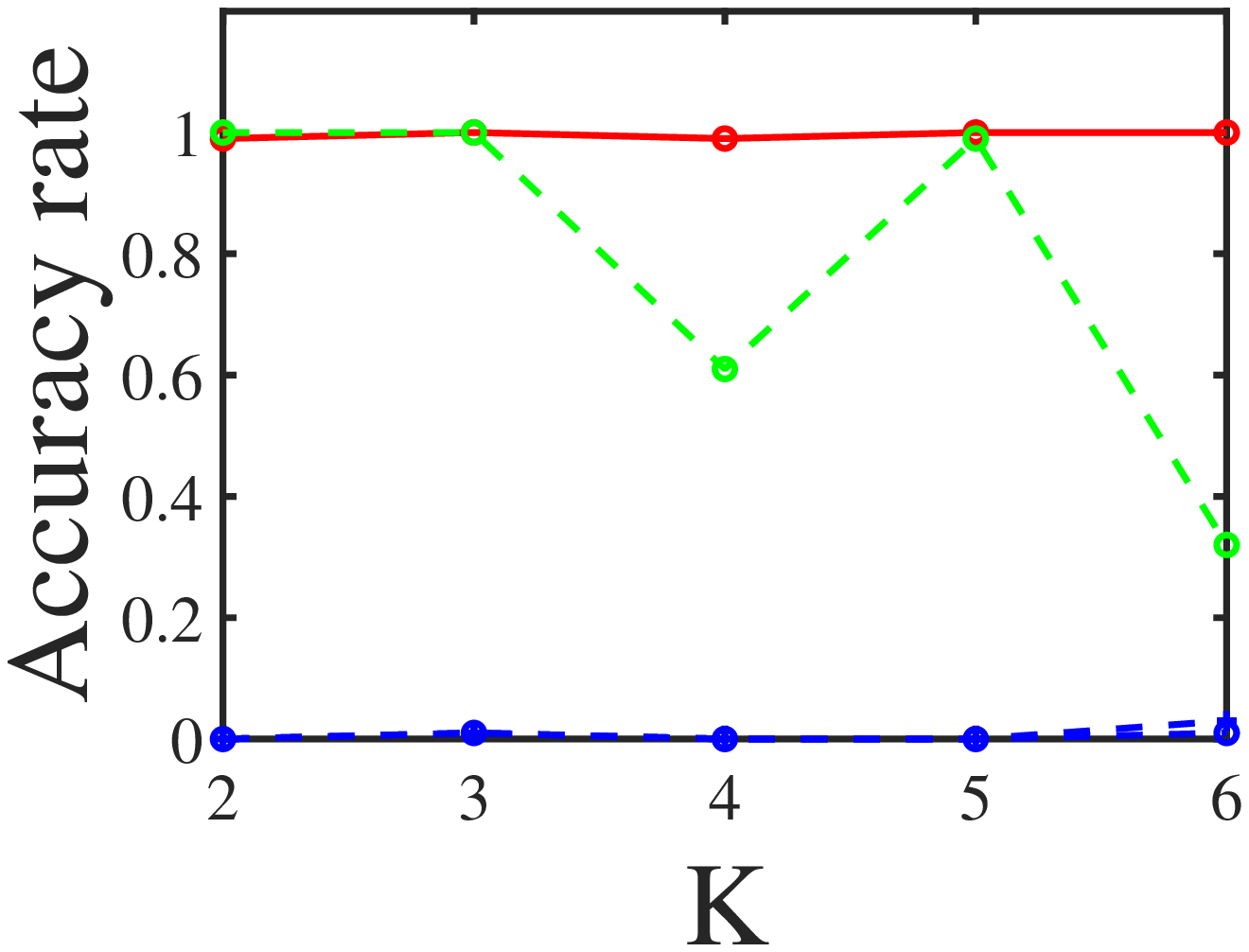}}
\subfigure[Experiment 5 (c)]{\includegraphics[width=0.24\textwidth]{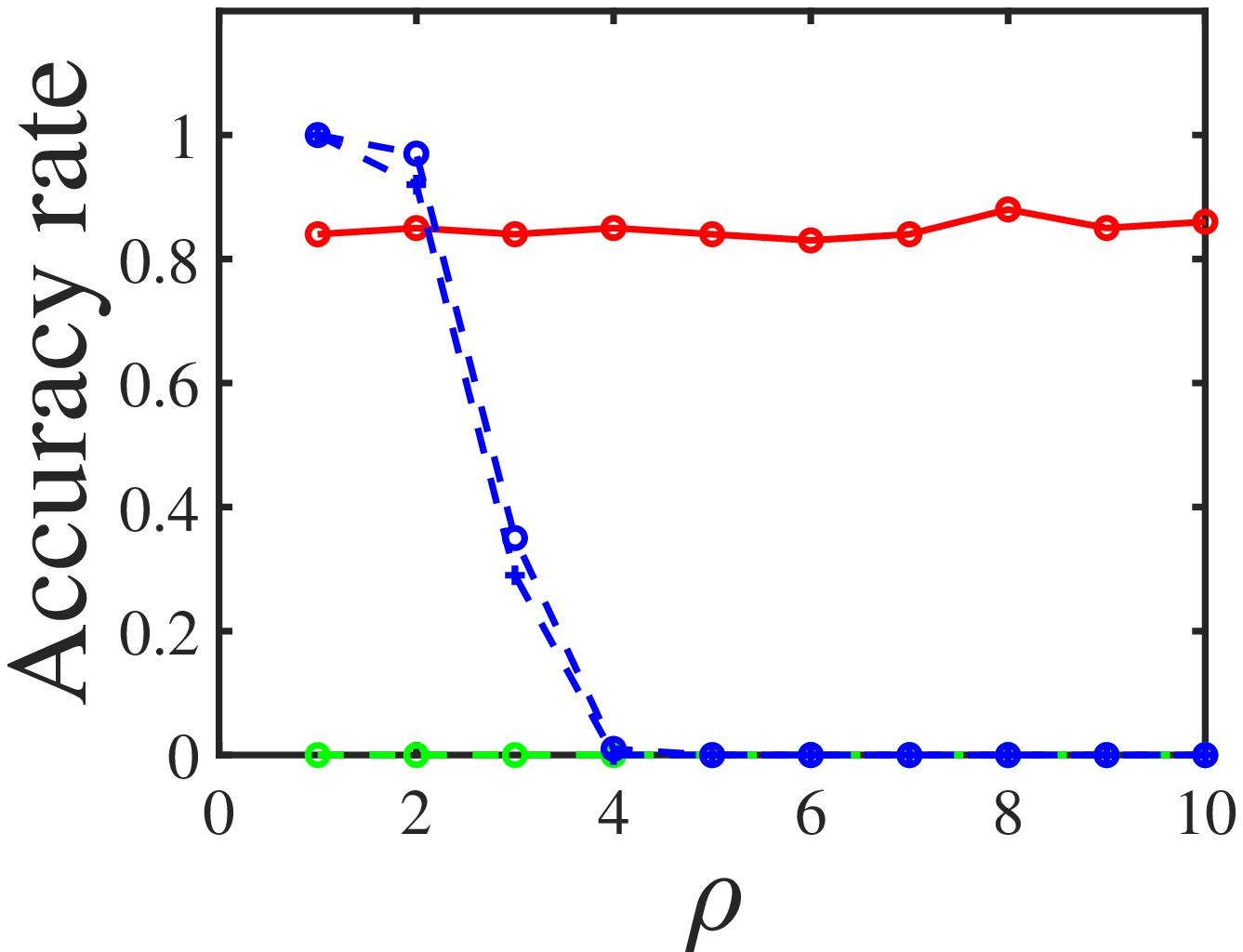}}
\subfigure[Experiment 5 (d)]{\includegraphics[width=0.24\textwidth]{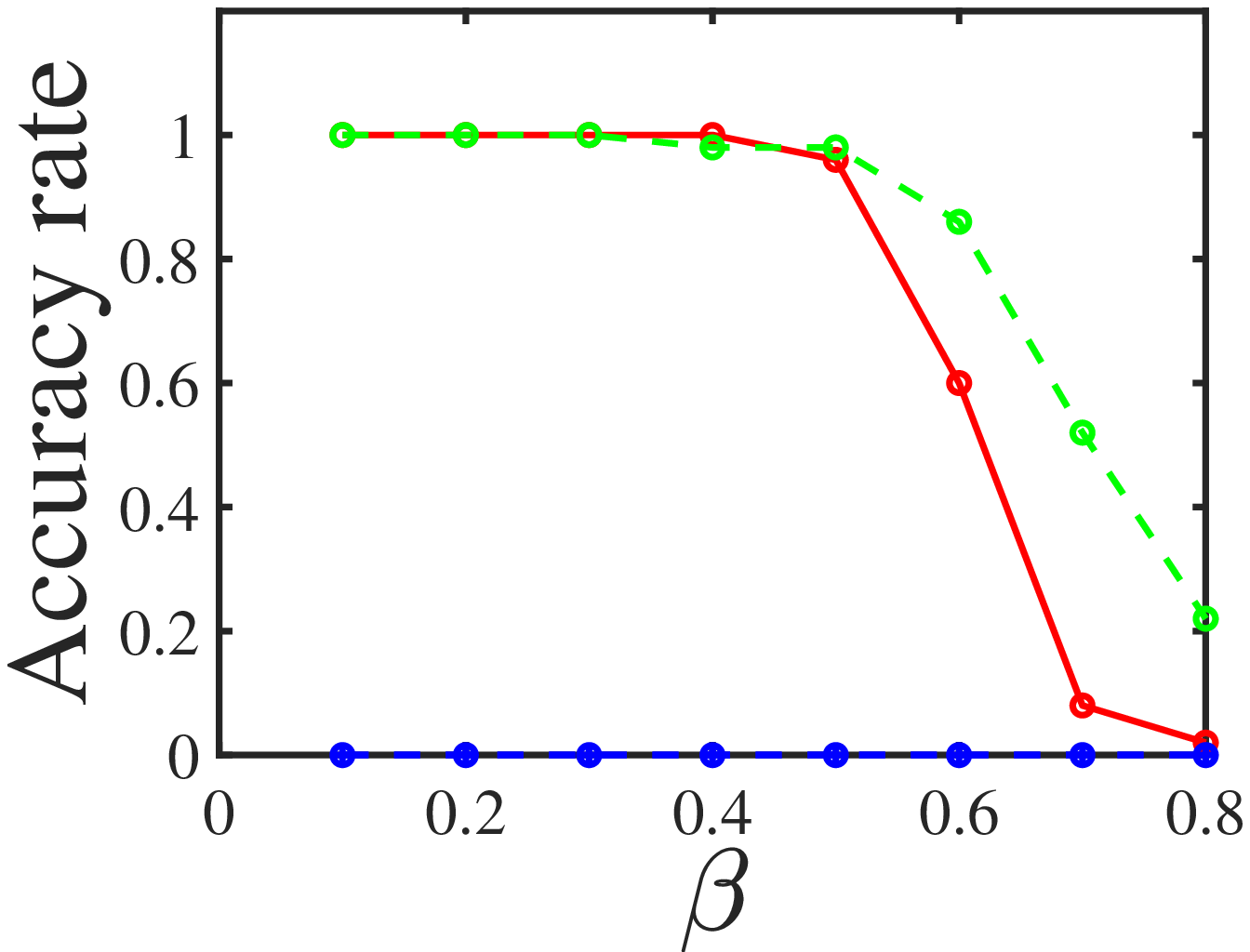}}
\caption{Exponential distribution.}
\label{S5} 
\end{figure}

\texttt{Experiment 5 (a): changing $\rho$.} Let $K=3$ and $P$ be the same as that of Experiment 1 (a). Let $\rho$ range in $\{1,2,\ldots, 10\}$.

\texttt{Experiment 5 (b): changing $K$.} Let $P$ be the same as Experiment 1 (b), $\rho=5$, and $K$ range in $\{2,3,\ldots, 6\}$.

\texttt{Experiment 5 (c): changing $\rho$ when $K=1$.} Let $K=1, P=1$, and $\rho$ range in $\{1, 2, \ldots, 10\}$.

\texttt{Experiment 5 (d): connectivity across communities.} Let $K=2, \rho= 5$, and $P$ be the same as Experiment 1 (d).

Figure \ref{S5} shows the Accuracy rate of Experiment 5. In general, we see that our nDFAwm estimates $K$ more accurately than its competitors except Experiment 5 (d) where ME performs slightly better than our nDFAwm. From panels (a) and (c) of Figure \ref{S5}, it is interesting to find that NB and BHac perform poorer as $\rho$ increases. Panels (b) and (d) of Figure \ref{S5} say that NB and BHac fail to work for Experiments 5 (b) and 5 (d).
\subsection{Normal distribution}
When $\mathcal{F}$ is Normal distribution such that $A_{ij}\sim \mathrm{Normal}(\Omega_{ij},\sigma^{2})$, i.e., $A_{ij}\in\mathbb{R}$ for $i,j\in[n]$, where $\Omega(i,j), \sigma^{2}$ are the expectation and variance terms of Normal distribution, respectively. By the property of Normal distribution, $\mathbb{E}[A_{ij}]=\Omega_{ij}$ satisfies Equation (\ref{ADCDFM}) and all entries of $P$ are real values. So, $\rho$'s range is $(0,+\infty)$ and $P$'s elements can be negative.
\begin{figure}
\centering
\subfigure[Experiment 6 (a)]{\includegraphics[width=0.24\textwidth]{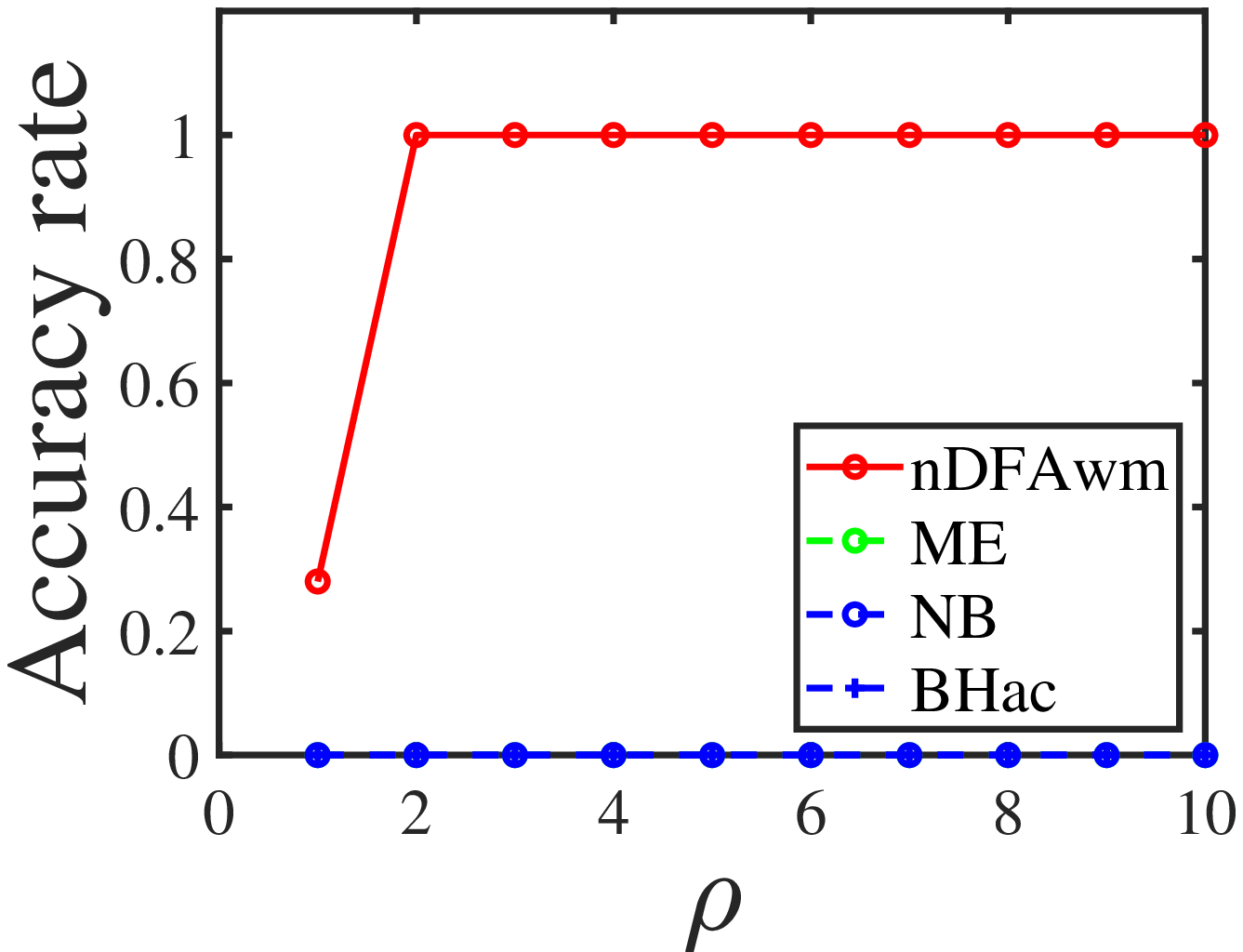}}
\subfigure[Experiment 6 (b)]{\includegraphics[width=0.24\textwidth]{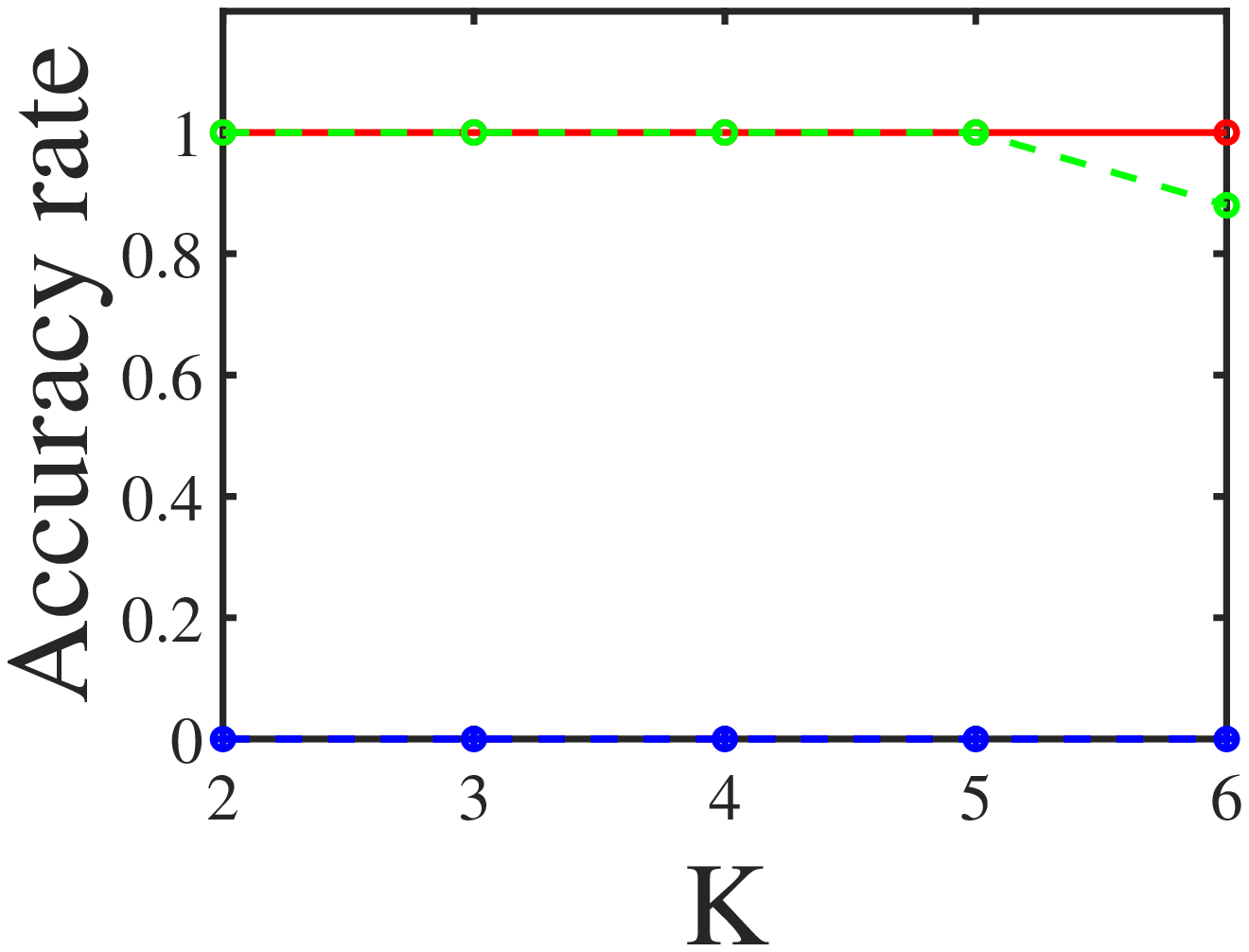}}
\subfigure[Experiment 6 (c)]{\includegraphics[width=0.24\textwidth]{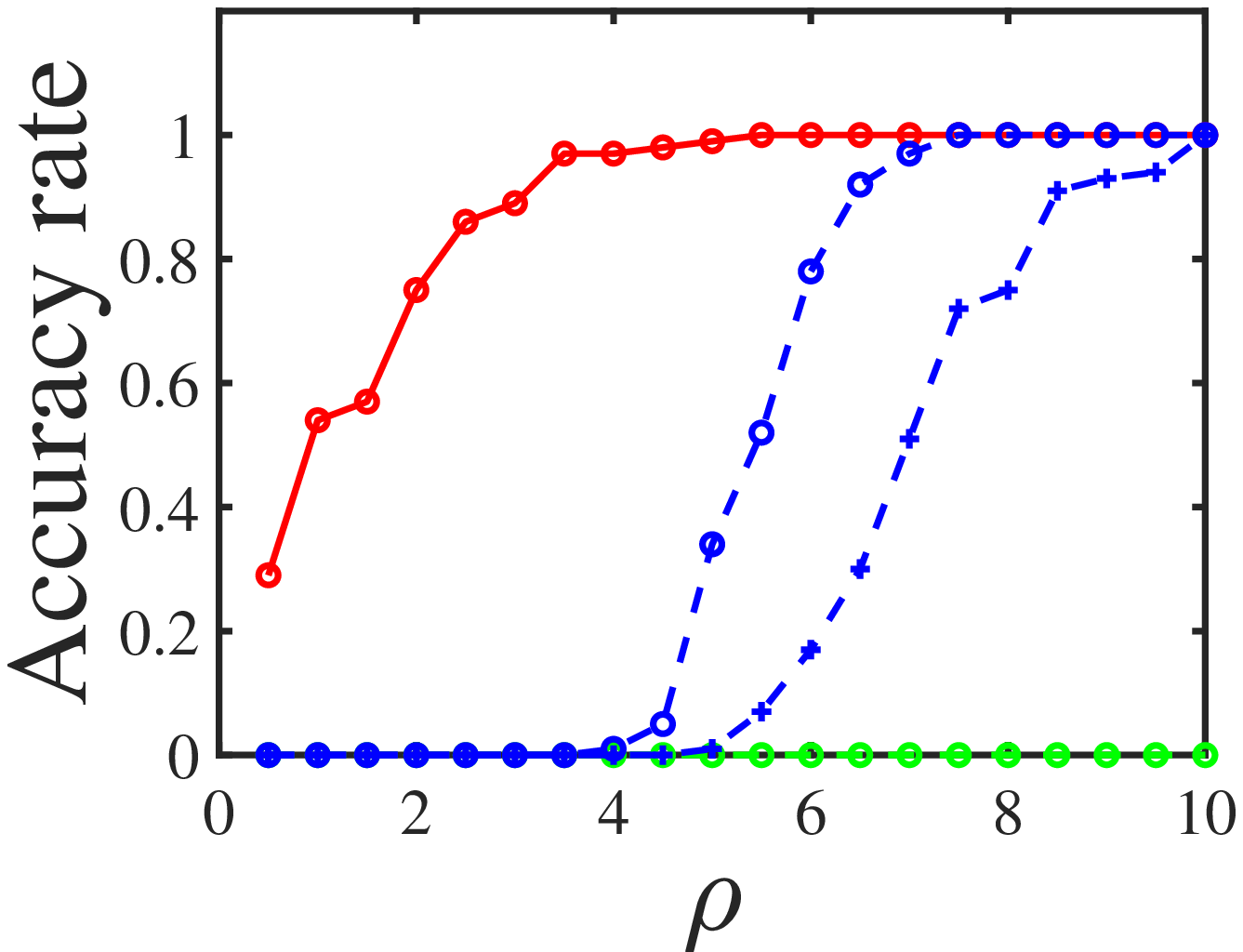}}
\subfigure[Experiment 6 (d)]{\includegraphics[width=0.24\textwidth]{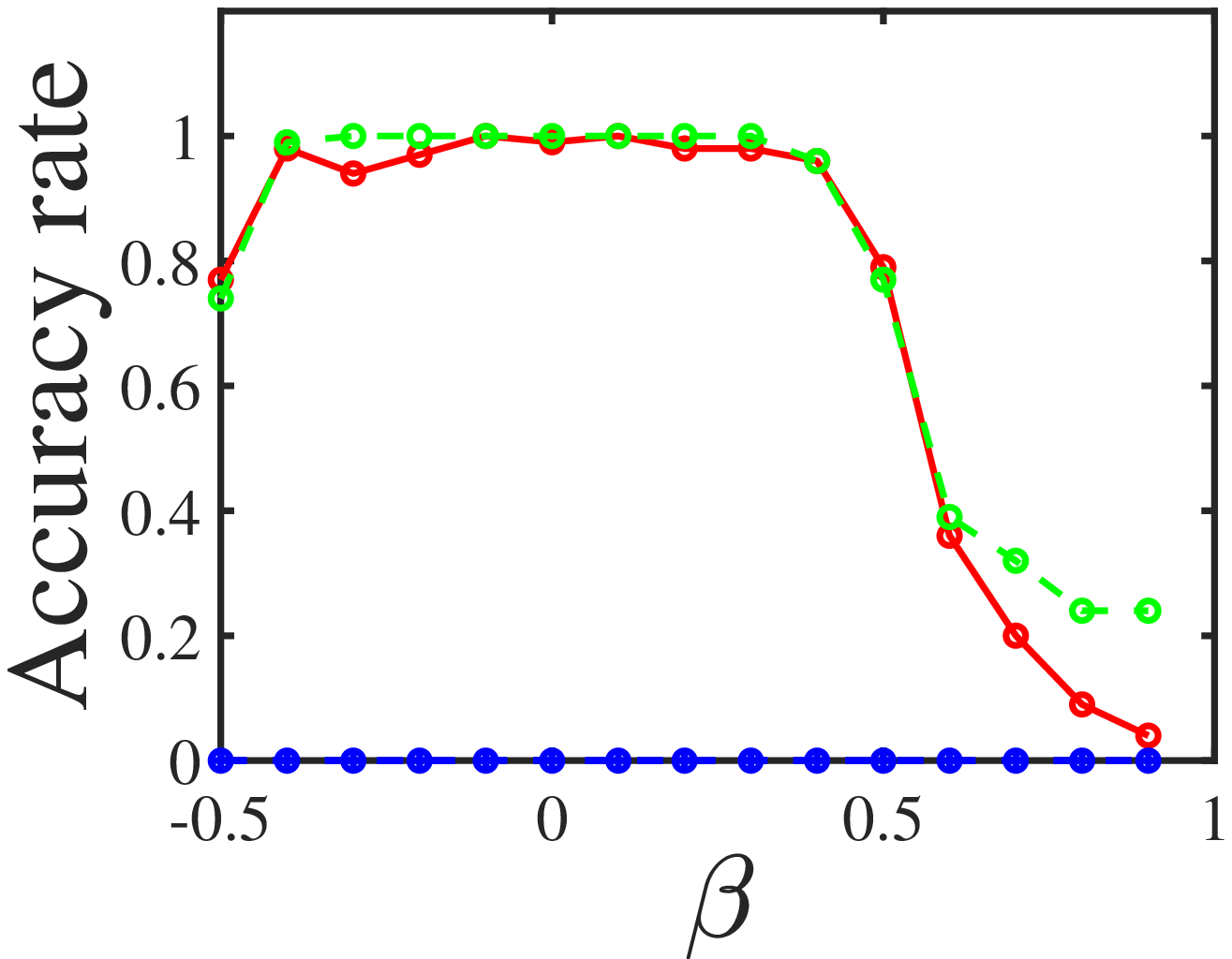}}
\caption{Normal distribution.}
\label{S6} 
\end{figure}

\texttt{Experiment 6 (a): changing $\rho$.} Let $K=3, \sigma^{2}=1$, and $P$ be
\[P=\begin{bmatrix}
    1&-0.2&-0.3\\
    -0.2&0.8&0.2\\
    -0.3&0.2&0.9\\
\end{bmatrix}.\]
Let $\rho$ range in $\{1, 2, \ldots, 10\}$.

\texttt{Experiment 6 (b): changing $K$.} Let $P$ be the same as Experiment 1 (b), $\sigma^{2}=1, \rho=3$, and $K$ range in $\{2,3,\ldots, 6\}$.

\texttt{Experiment 6 (c): changing $\rho$ when $K=1$.} Let $K=1, \sigma^{2}=1, P=1$, and $\rho$ range in $\{0.5, 1, \ldots, 10\}$.

\texttt{Experiment 6 (d): connectivity across communities.} Let $K=2, \sigma^{2}=1, \rho=2$, $P$'s diagonal entries be 1, $P$'s off-diagonal entries be $\beta$, and $\beta$ range in $\{-0.5, -0.4, \ldots, 0.9\}$.

Figure \ref{S6} shows the Accuracy rate of Experiment 6. In general, we see that our nDFAwm outperforms its competitors except for Experiment 6 (d) where it performs similarly to ME. From panels (a), (b), and (d) of Figure \ref{S6}, we see that NB and BHac fail to work. Panel (c) of Figure \ref{S6} says that though NB and BHac perform poorer than our nDFAwm, they provide more accurate estimations as $\rho$ increases for Experiment 6 (c).
\subsection{Laplace distribution}
When $\mathcal{F}$ is Laplace distribution such that $A_{ij}\sim \mathrm{Laplace}(\Omega_{ij},\frac{\sigma^{2}}{2})$, i.e., $A_{ij}\in\mathbb{R}$ for $i,j\in[n]$, where $\Omega(i,j), \sigma^{2}$ are the expectation and variance terms of Laplace distribution, respectively. Similar to Normal distribution, $\mathbb{E}[A_{ij}]=\Omega_{ij}$ satisfies Equation (\ref{ADCDFM}), all elements of $P$ are real values, and $\rho$'s range is $(0,+\infty)$.
\begin{figure}
\centering
\subfigure[Experiment 7 (a)]{\includegraphics[width=0.24\textwidth]{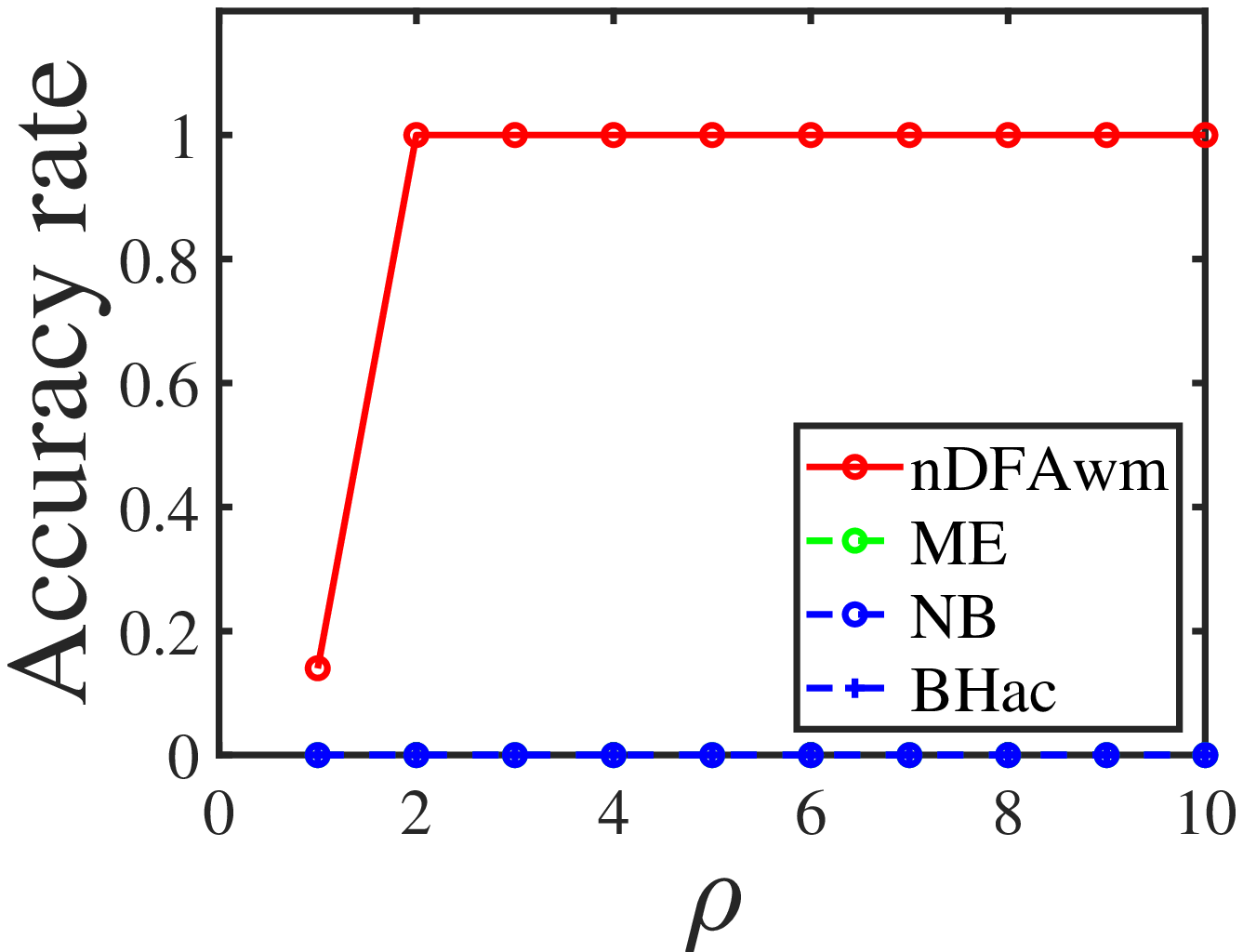}}
\subfigure[Experiment 7 (b)]{\includegraphics[width=0.24\textwidth]{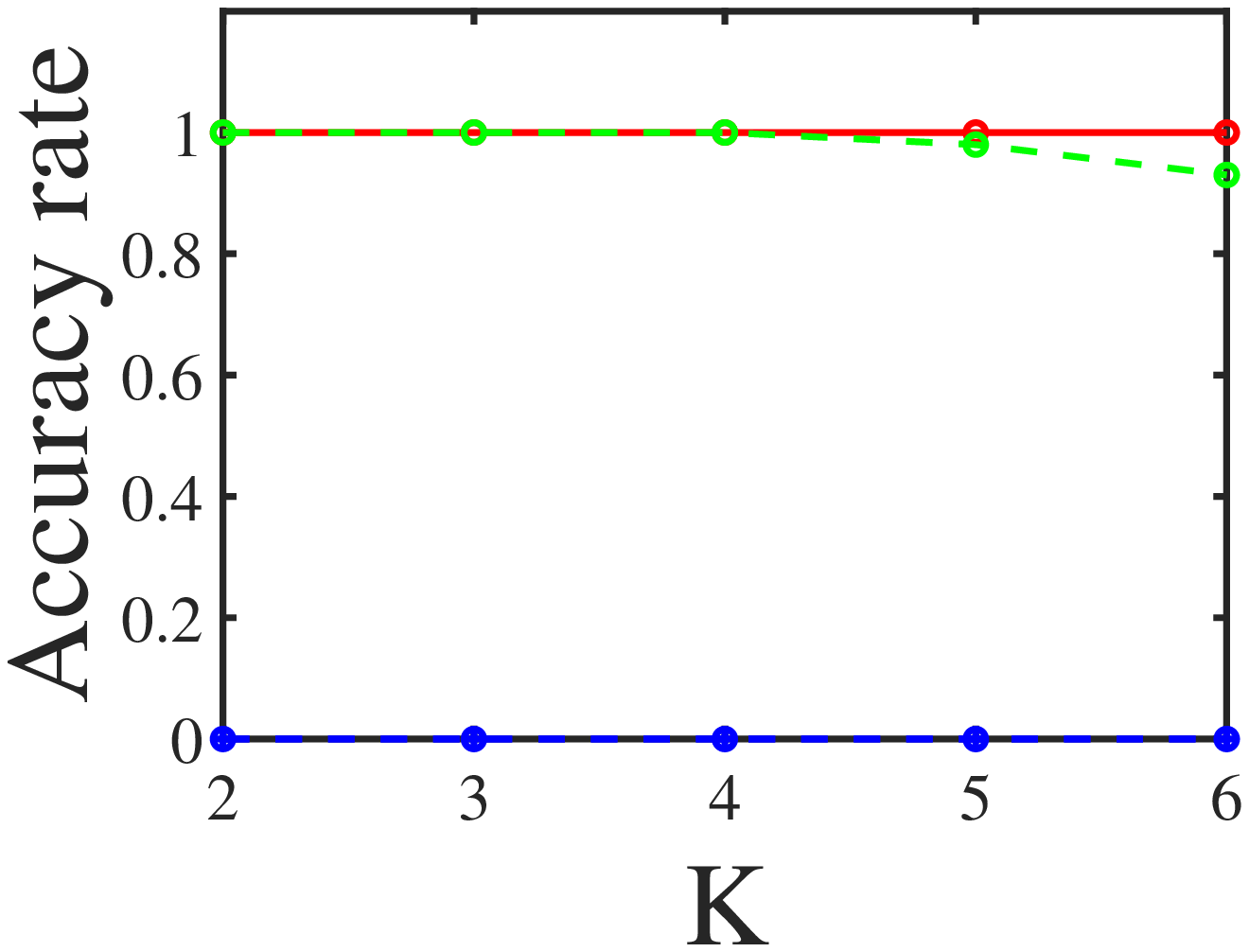}}
\subfigure[Experiment 7 (c)]{\includegraphics[width=0.24\textwidth]{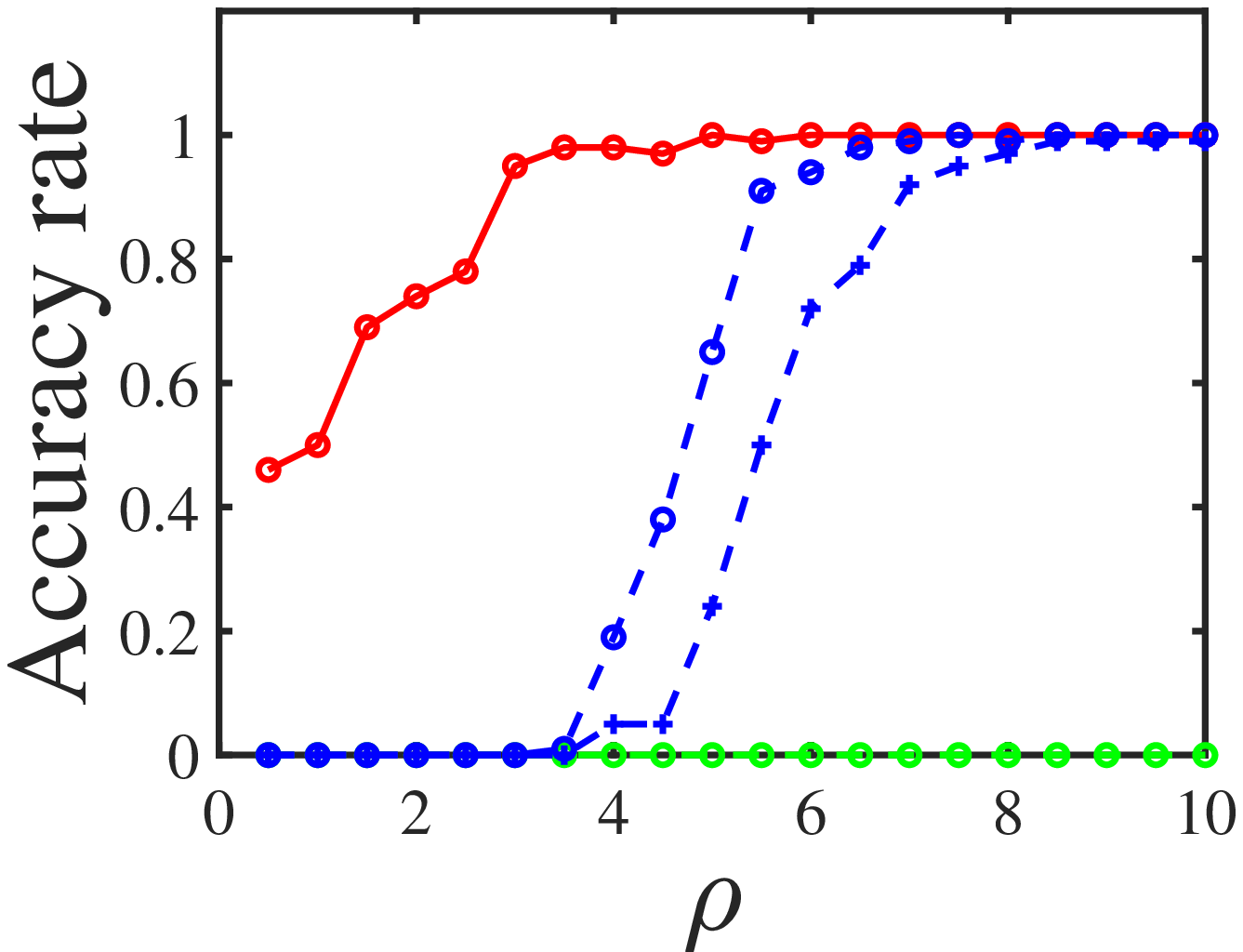}}
\subfigure[Experiment 7 (d)]{\includegraphics[width=0.24\textwidth]{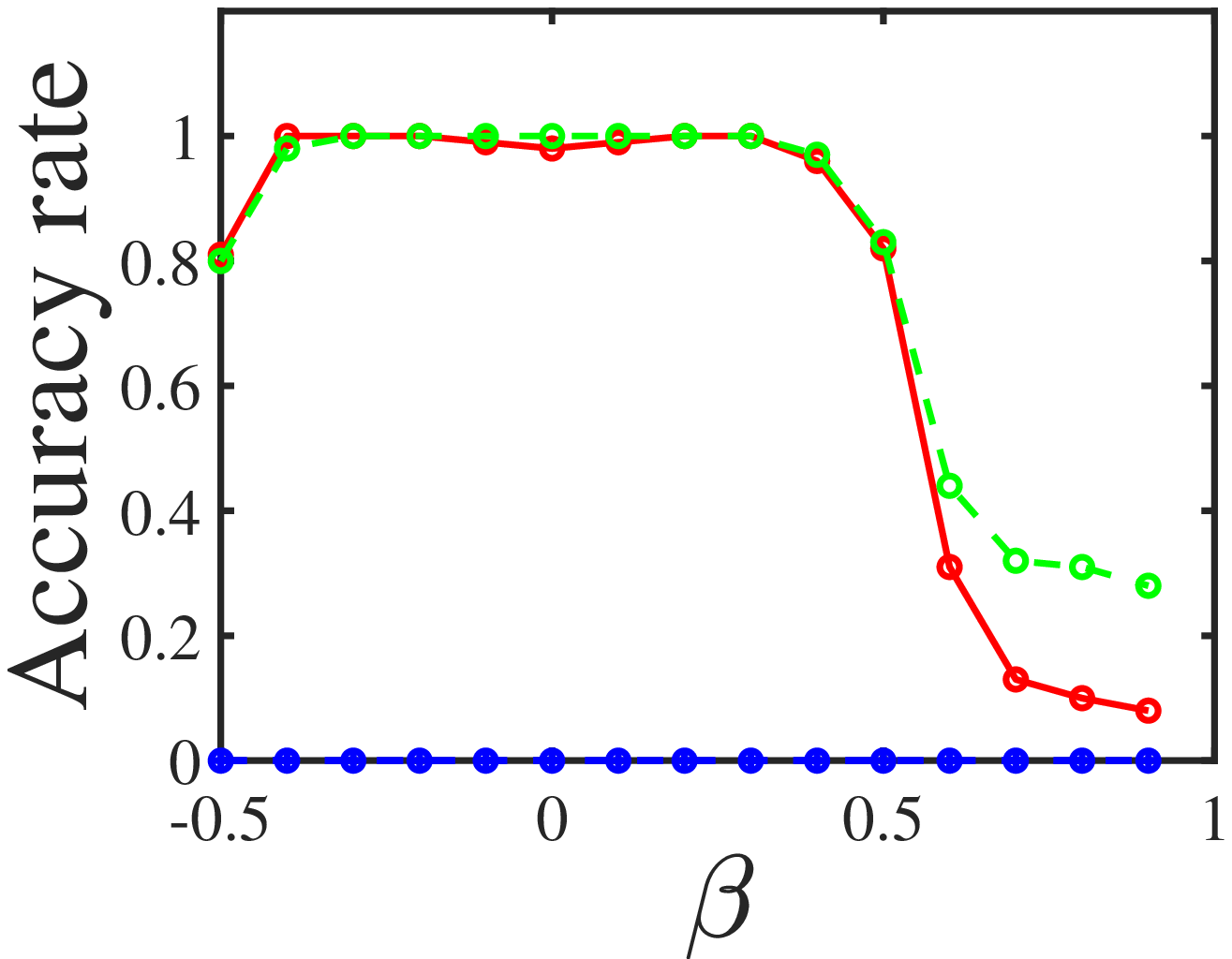}}
\caption{Laplace distribution.}
\label{S7} 
\end{figure}

\texttt{Experiment 7 (a): changing $\rho$.} Let $K=3, \sigma^{2}=1$, $P$ be the same as Experiment 6 (a), and $\rho$ range in $\{1, 2, \ldots, 10\}$.

\texttt{Experiment 7 (b): changing $K$.} Let $P$ be the same as Experiment 1 (b), $\sigma^{2}=1, \rho=3$, and $K$ range in $\{2,3,\ldots, 6\}$.

\texttt{Experiment 7 (c): changing $\rho$ when $K=1$.} Let $K=1, \sigma^{2}=1, P=1$, and $\rho$ range in $\{0.5, 1, \ldots, 10\}$.

\texttt{Experiment 7 (d): connectivity across communities.} Let $K=2, \sigma^{2}=1, \rho=2$, $P$'s diagonal entries be 1, $P$'s off-diagonal entries be $\beta$, and $\beta$ range in $\{-0.5, -0.4, \ldots, 0.9\}$.

Figure \ref{S7} displays the Accuracy rate of Experiment 7. The numerical results are similar to that of Experiment 6 and we omit the analysis here.
\subsection{Uniform distribution}
When $\mathcal{F}$ is Uniform distribution such that $A_{ij}\sim \mathrm{Uniform}(0,\Omega_{ij})$, i.e., $A_{ij}\in(0,\mathrm{max}_{i,j\in[n]}\Omega_{ij})$. For this case, $\mathbb{E}[A_{ij}]=\Omega_{ij}$ satisfies Equation (\ref{ADCDFM}), all elements of $P$ are nonnegative, and $\rho$'s range is $(0,+\infty)$.
\begin{figure}
\centering
\subfigure[Experiment 8 (a)]{\includegraphics[width=0.24\textwidth]{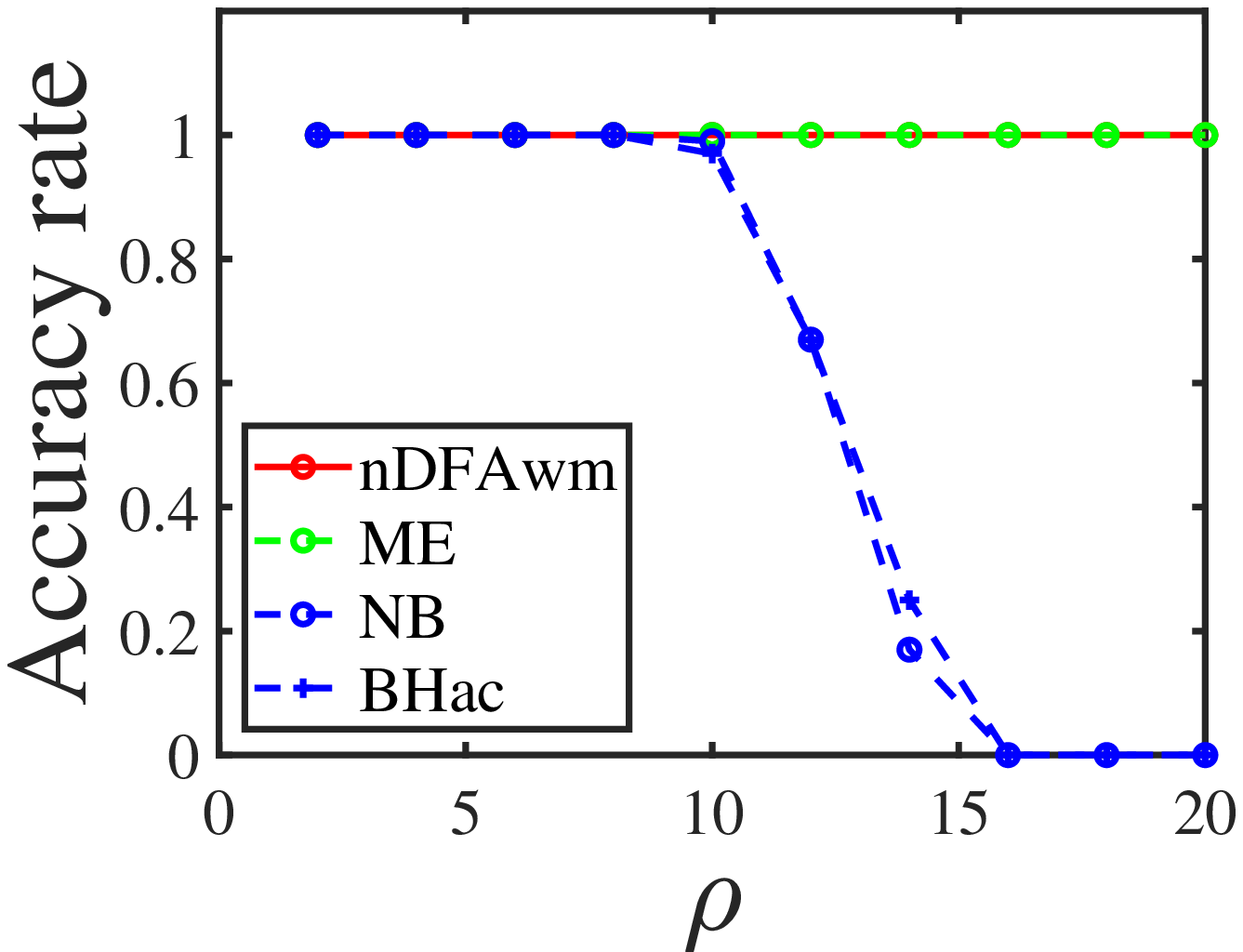}}
\subfigure[Experiment 8 (b)]{\includegraphics[width=0.24\textwidth]{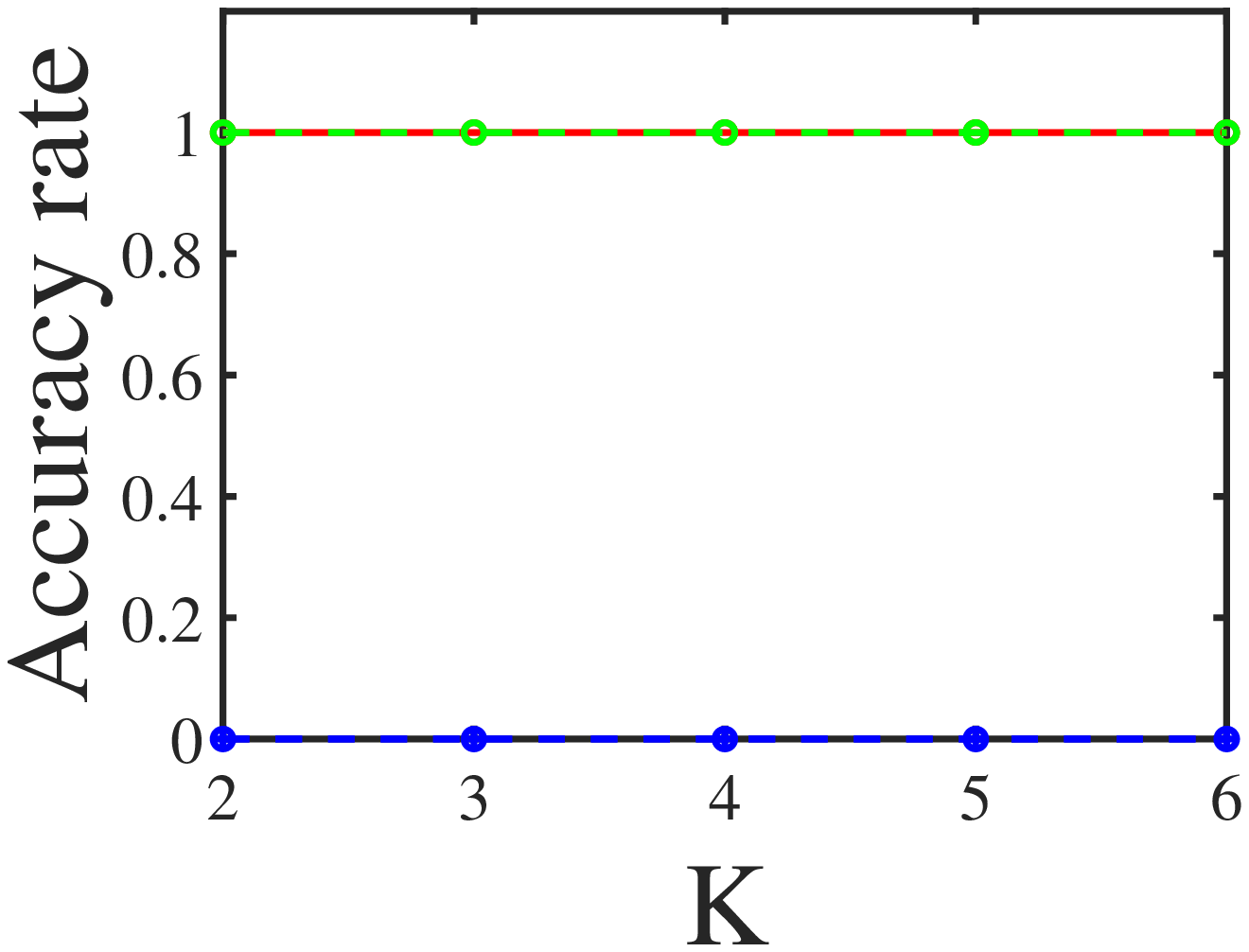}}
\subfigure[Experiment 8 (c)]{\includegraphics[width=0.24\textwidth]{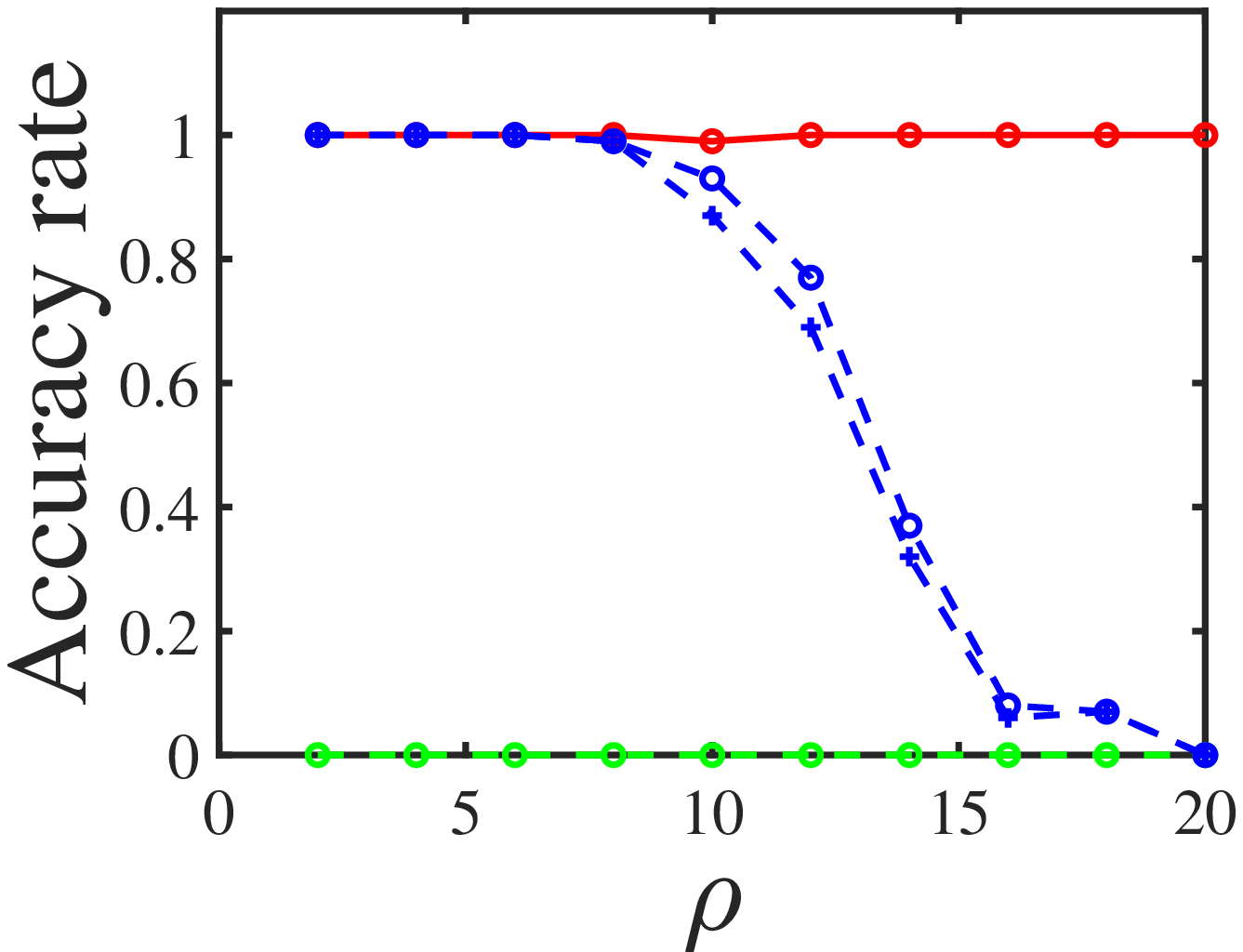}}
\subfigure[Experiment 8 (d)]{\includegraphics[width=0.24\textwidth]{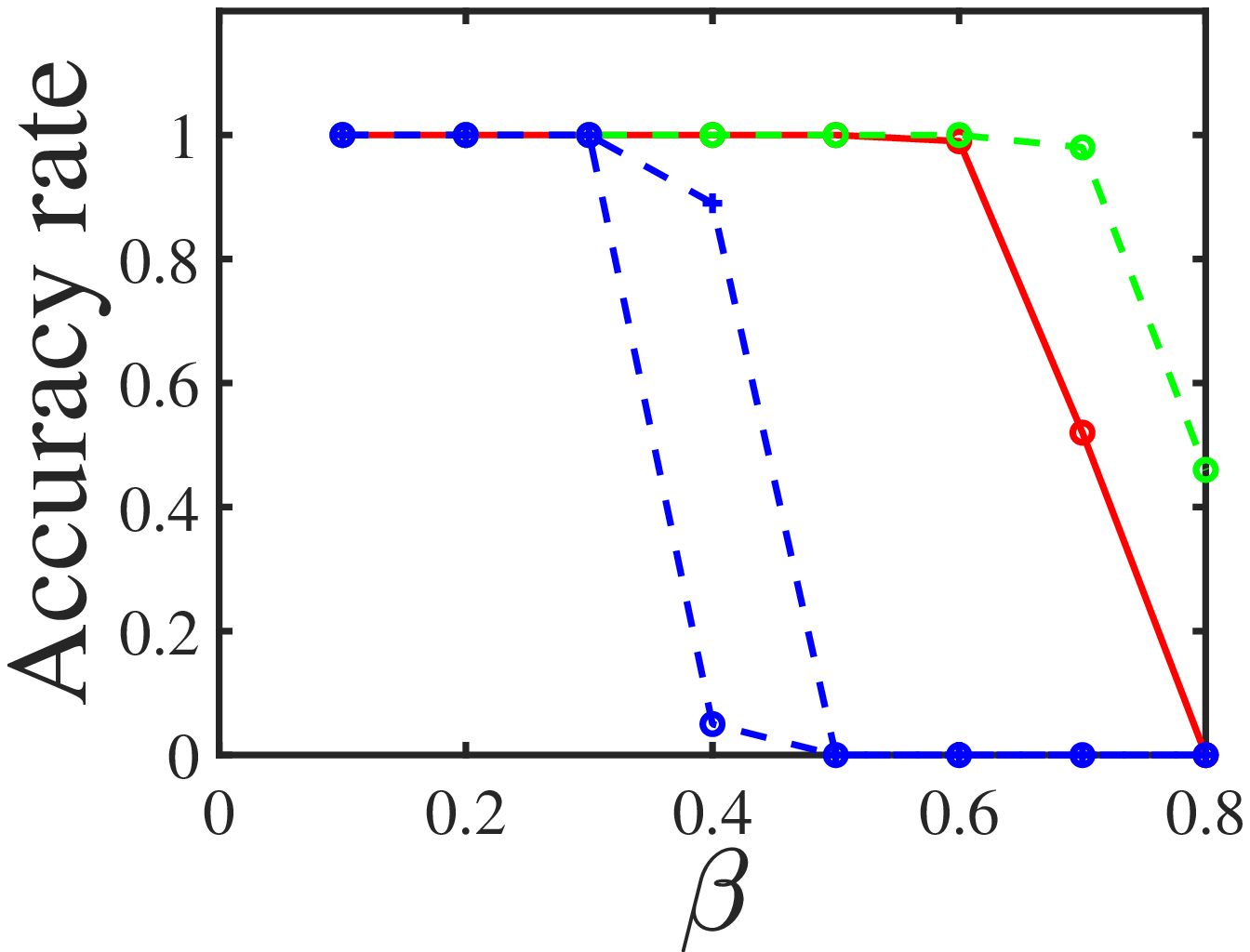}}
\caption{Uniform distribution.}
\label{S8} 
\end{figure}

\texttt{Experiment 8 (a): changing $\rho$.} Let $K=3$, $P$ be the same as Experiment 1 (a), and $\rho$ range in $\{2, 4, \ldots, 20\}$.

\texttt{Experiment 8 (b): changing $K$.} Let $P$ be the same as Experiment 1 (b), $\rho=0.3$, and $K$ range in $\{2,3,\ldots, 6\}$.

\texttt{Experiment 8 (c): changing $\rho$ when $K=1$.} Let $K=1, P=1$, and $\rho$ range in $\{2, 4, \ldots, 20\}$.

\texttt{Experiment 8 (d): connectivity across communities.} Let $K=2, \rho=1$, and $P$ be the same as Experiment 1 (d).

Figure \ref{S8} displays the Accuracy rate of Experiment 8. We see that our approach nDFAwm outperforms its competitors in all cases except for Experiment 8 (d) where it performs slightly poorer than ME. For ME method, it enjoys similar performances as our nDFAwm for Experiments 8 (a), 8(b), and 8 (d) while it fails to estimate the number of clusters when the true $K$ is 1. For NB and BHac, they perform poorer as $\rho$ increases for Experiments 8 (a), 8 (c), and 8 (d). Meanwhile, NB and BHac fail to work for Experiment 8 (b).
\subsection{Signed networks}
Let $\mathbb{P}(A_{ij}=1)=\frac{1+\Omega_{ij}}{2}$ and $\mathbb{P}(A_{ij}=-1)=\frac{1-\Omega_{ij}}{2}$ such that $A$ is the adjacency matrix of a signed network. For this case, $\mathbb{E}[A_{ij}]=\Omega_{ij}$ satisfies Equation (\ref{ADCDFM}), all elements of $P$ are real values, and $\rho$'s range is $(0,1]$. For signed networks, we let $n=100K$, each node belong to one of the $K$ communities with equal probability, and $\theta_{i}=\sqrt{\rho}$ for $i\in[n]$.
\begin{figure}
\centering
\subfigure[Experiment 9 (a)]{\includegraphics[width=0.24\textwidth]{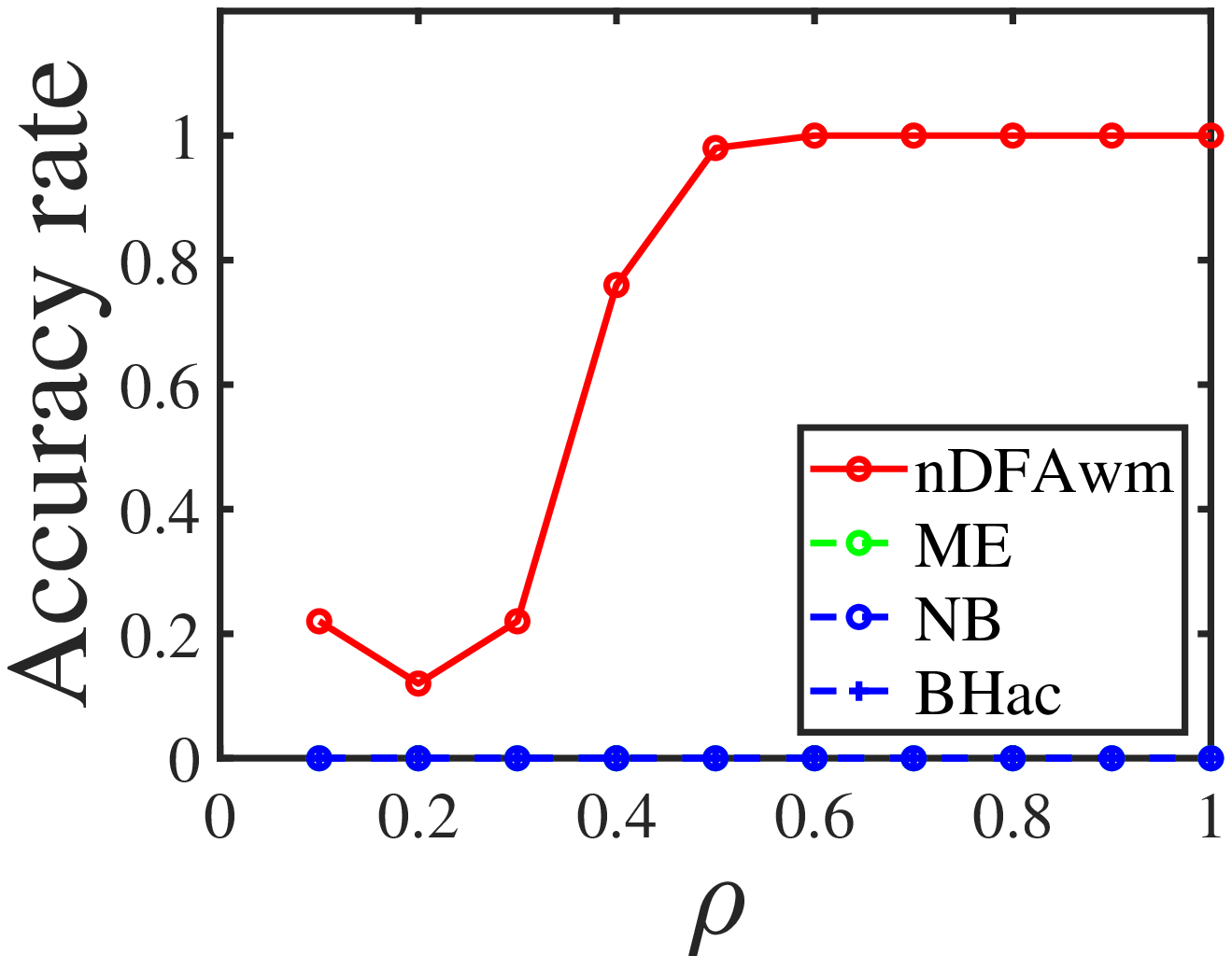}}
\subfigure[Experiment 9 (b)]{\includegraphics[width=0.24\textwidth]{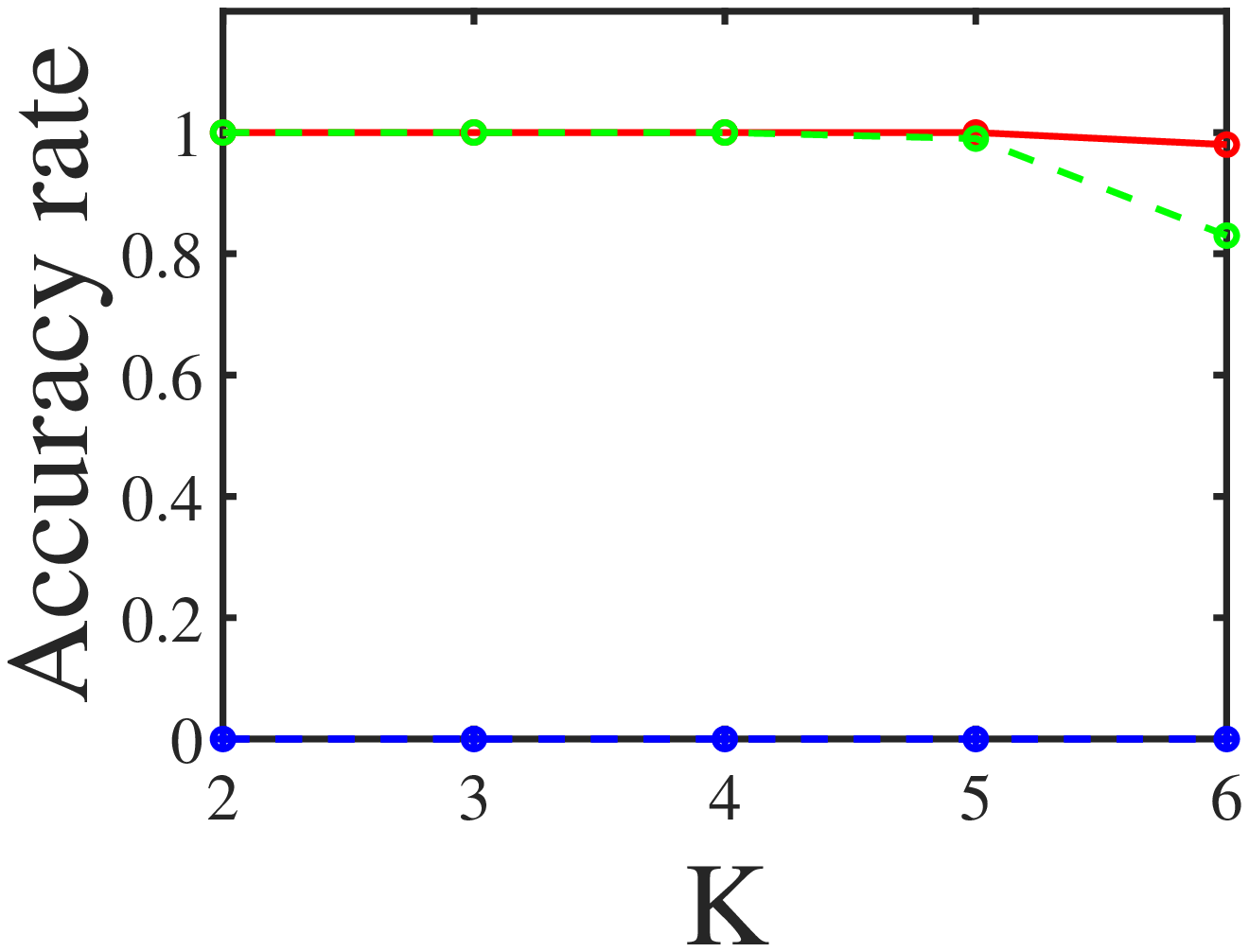}}
\subfigure[Experiment 9 (c)]{\includegraphics[width=0.24\textwidth]{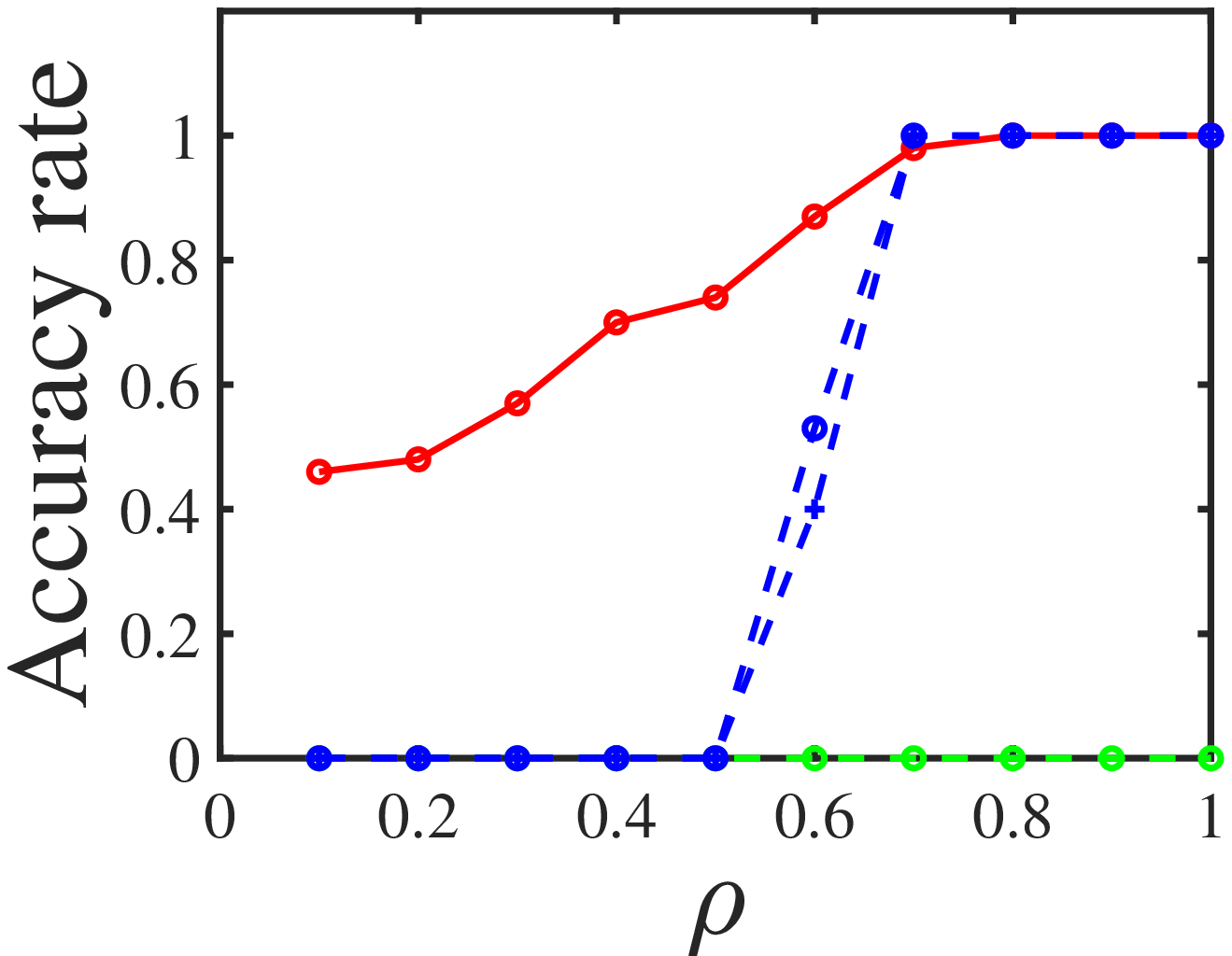}}
\subfigure[Experiment 9 (d)]{\includegraphics[width=0.24\textwidth]{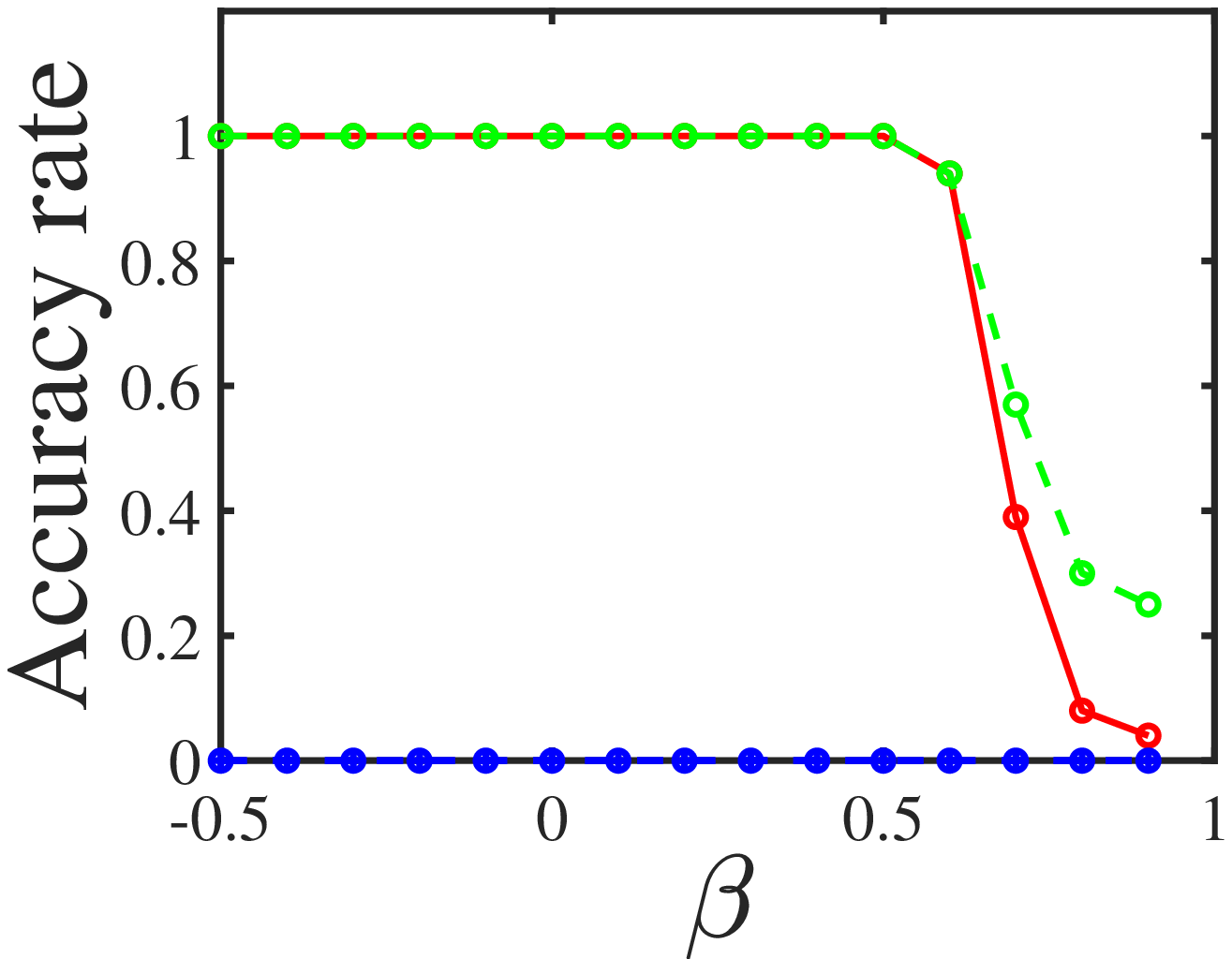}}
\caption{Signed networks.}
\label{S9} 
\end{figure}

\texttt{Experiment 9 (a): changing $\rho$.} Let $K=3$, $P$ be the same as Experiment 6 (a), and $\rho$ range in $\{0.1, 0.2, \ldots, 1\}$.

\texttt{Experiment 9 (b): changing $K$.} Let $P$ be the same as Experiment 1 (b), $\rho=0.5$, and $K$ range in $\{2,3,\ldots, 6\}$.

\texttt{Experiment 9 (c): changing $\rho$ when $K=1$.} Let $K=1, P=1$, and $\rho$ range in $\{0.1, 0.2, \ldots, 1\}$.

\texttt{Experiment 9 (d): connectivity across communities.} Let $K=2, \rho=0.5$, $P$'s diagonal entries be 1, $P$'s off-diagonal entries be $\beta$, and $\beta$ range in $\{-0.5, -0.4, \ldots, 0.9\}$.

Figure \ref{S9} displays the Accuracy rate of Experiment 9. We see that our approach nDFAwm provides a more accurate estimation of the number of clusters than its competitors except Experiment 9 (d) where it performs similarly to ME. For ME, it fails to work in Experiments 9 (a) and 9 (c). For NB and BHac, they fail to estimate $K$ except for Experiment 9 (c) where they have better estimations as $\rho$ increases.
\section{Real-world networks}
For real-world networks, we consider eight data sets in Table \ref{realdata}. The ground truth numbers of communities of these eight networks are known and they provide a reasonable baseline to compare estimators. The Karate club (weighted) network is a weighted network with nonnegative edge weights, the Gahuku-Gama subtribes is a signed network, the Slovene Parliamentary Party network is a weighted network with positive and negative edge weights, and the other five data sets are unweighted. The Karate club (weighted) network can be downloaded from \url{http://vlado.fmf.uni-lj.si/pub/networks/data/ucinet/ucidata.htm#kazalo} and it is the weighted version of the classical Karate club network. The Gahuku-Gama subtribes network can be downloaded from \url{http://konect.cc/networks/ucidata-gama/} and its ground truth of node labels can be found in Figure 9 (b) of \cite{yang2007community}. The Slovene Parliamentary Party network can be downloaded from \url{http://vlado.fmf.uni-lj.si/pub/networks/data/soc/Samo/Stranke94.htm}. The other five data sets with ground truth of node labels can be downloaded from \url{http://www-personal.umich.edu/~mejn/netdata/}. In particular, for the Dolphins network, as analyzed in \cite{liu2016discovering}, both $K=2$ or $K=4$ are reasonable.
\begin{figure}
\centering
\subfigure[Karate club (weighted)]{\includegraphics[width=0.24\textwidth]{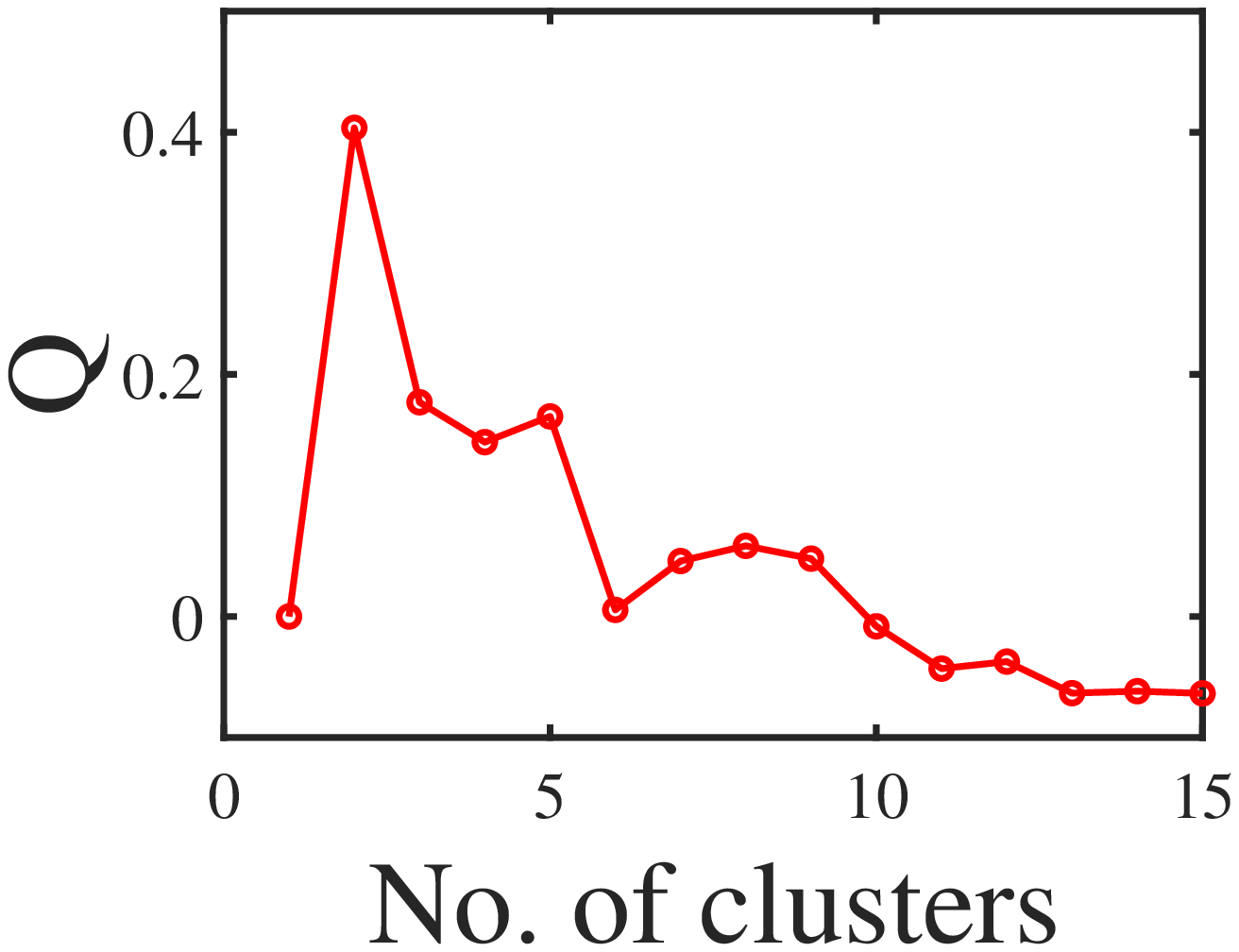}}
\subfigure[Gahuku-Gama subtribes]{\includegraphics[width=0.24\textwidth]{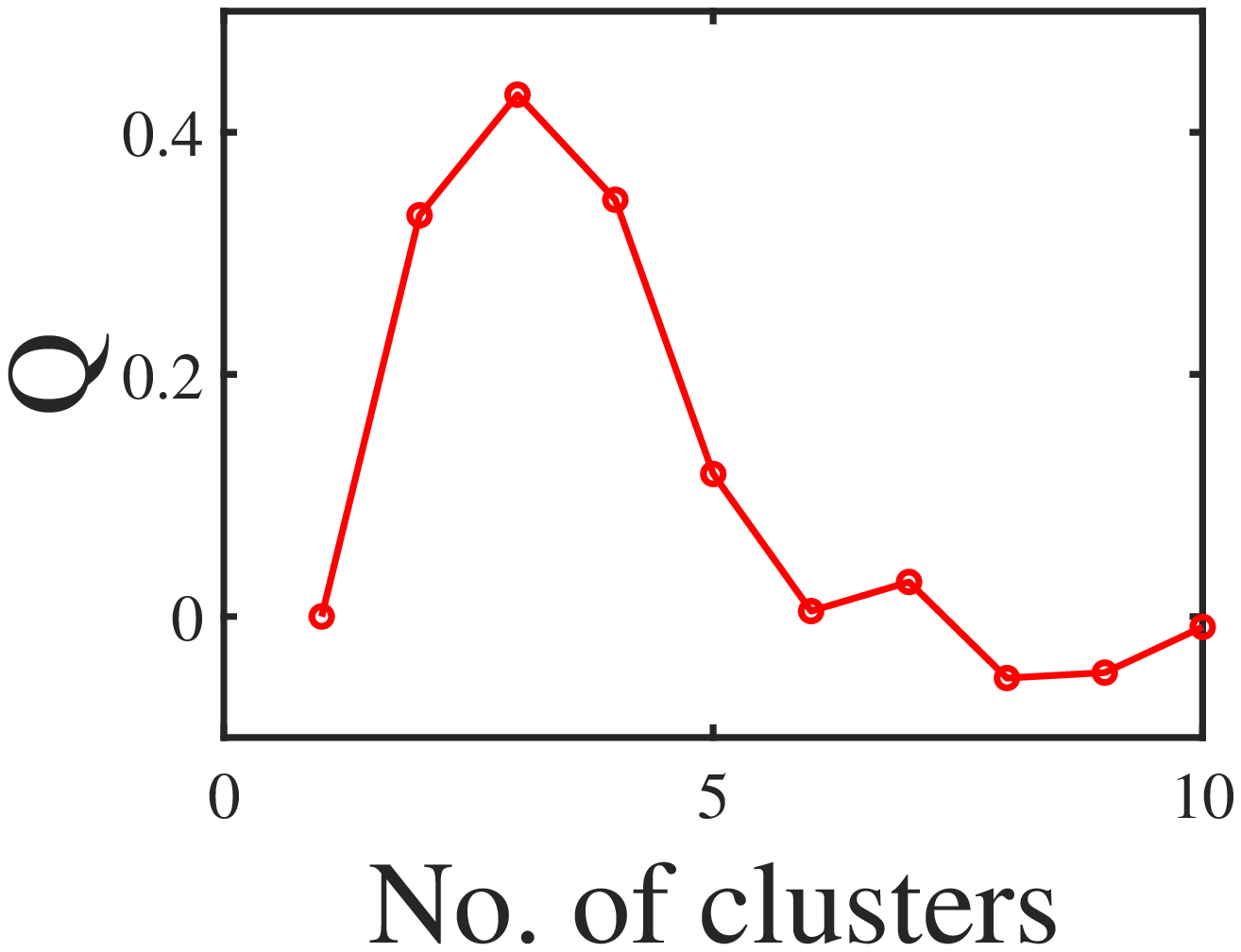}}
\subfigure[Slovene Parliamentary Party network]{\includegraphics[width=0.24\textwidth]{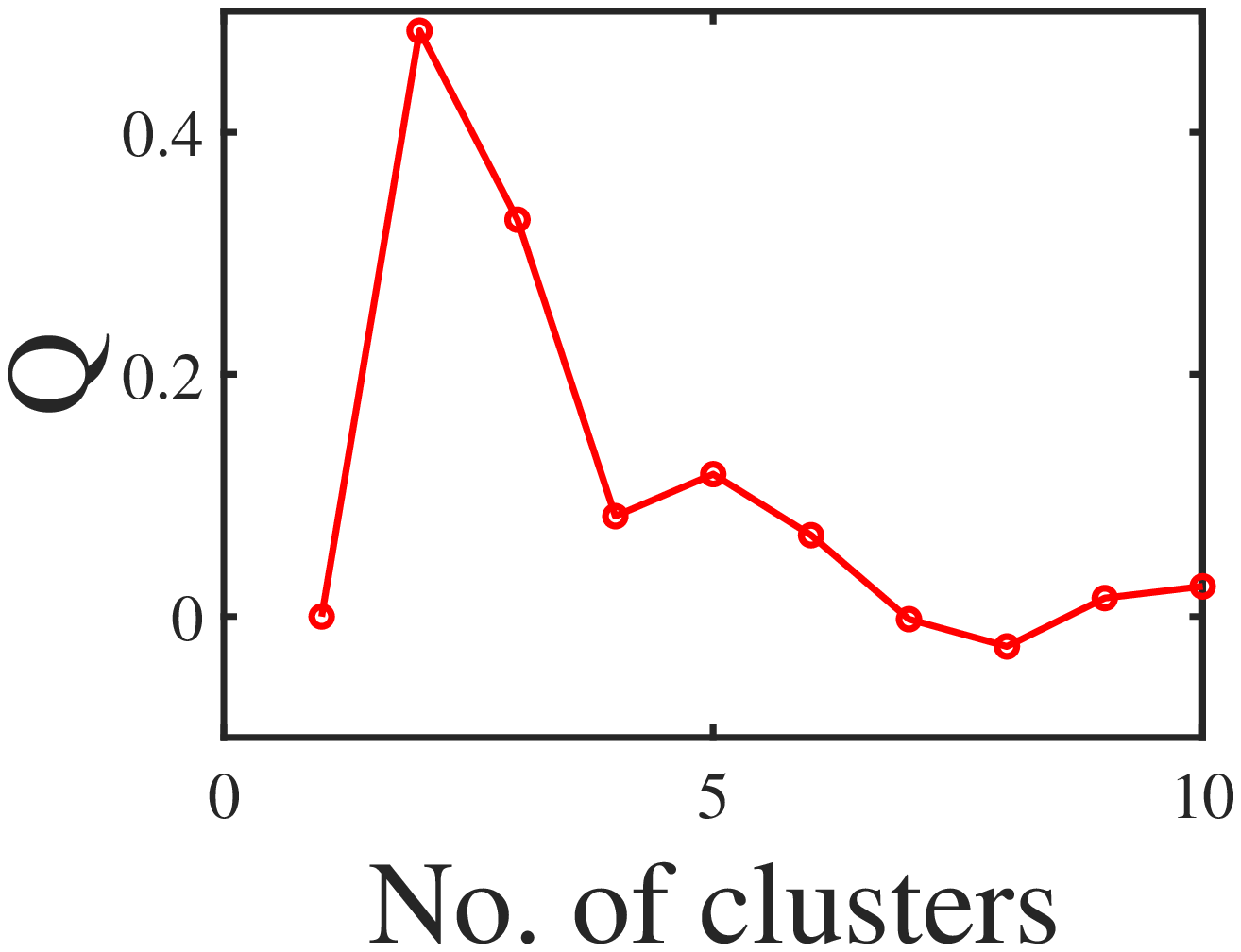}}
\subfigure[Dolphins]{\includegraphics[width=0.24\textwidth]{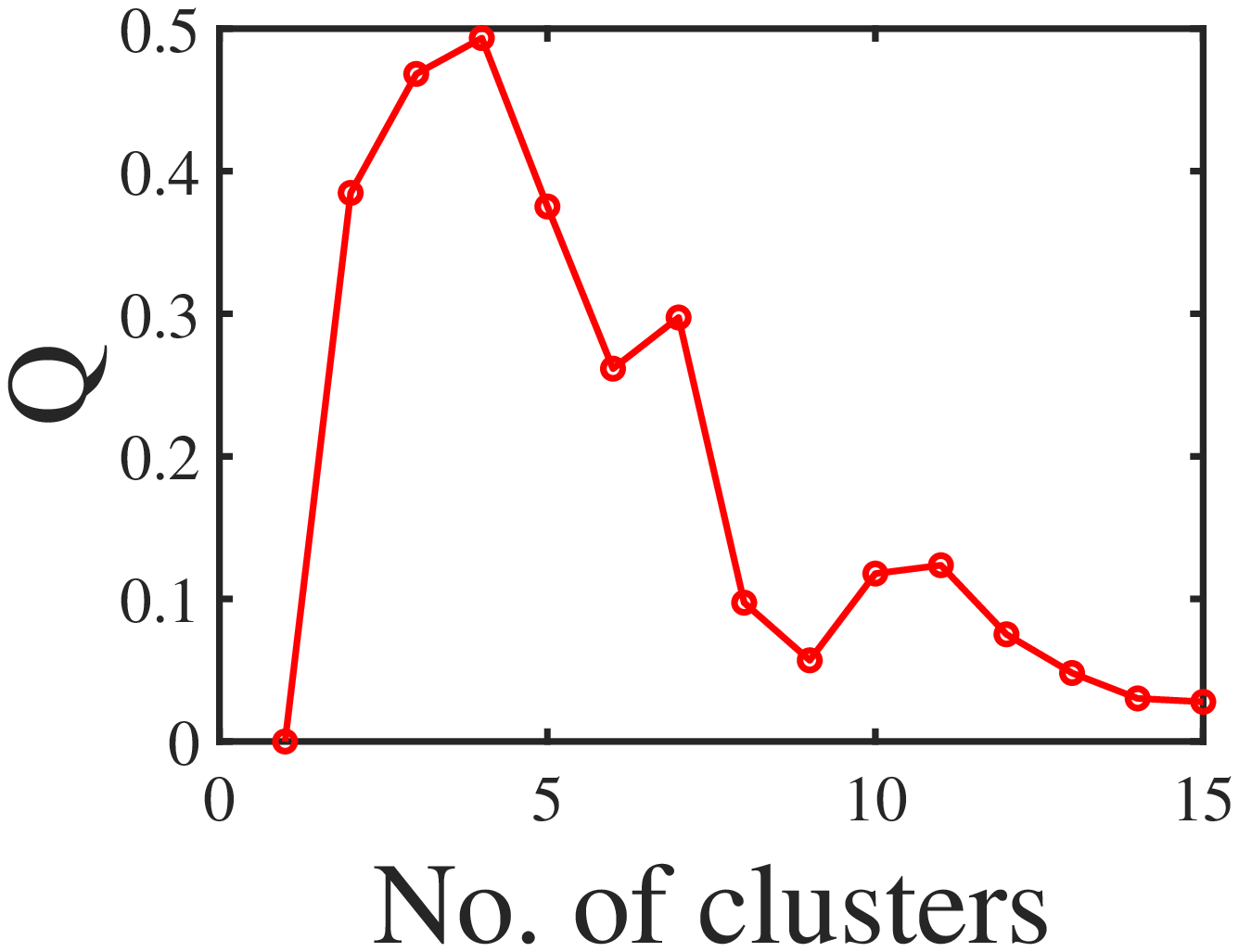}}
\subfigure[College football]{\includegraphics[width=0.24\textwidth]{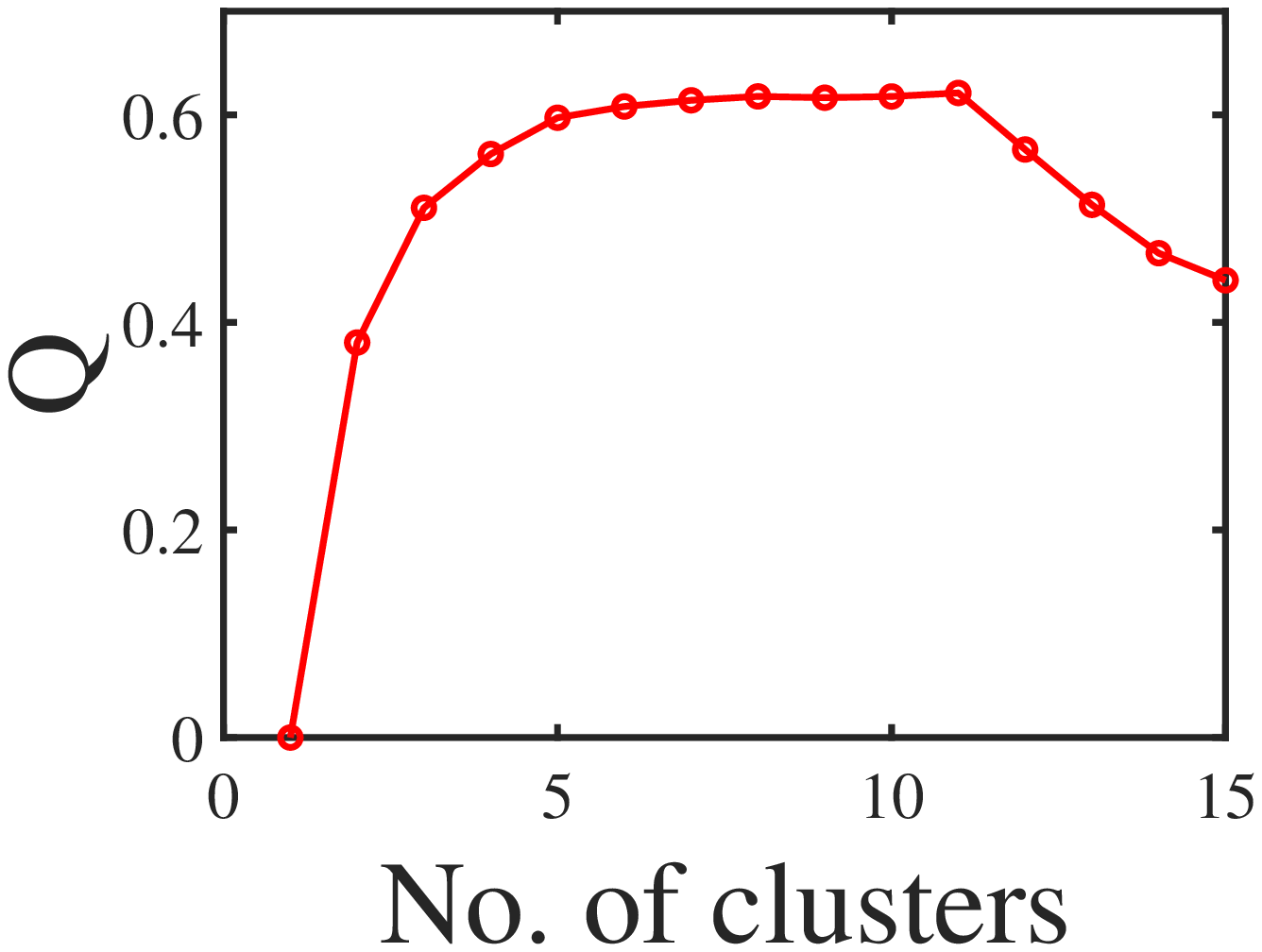}}
\subfigure[Karate club]{\includegraphics[width=0.24\textwidth]{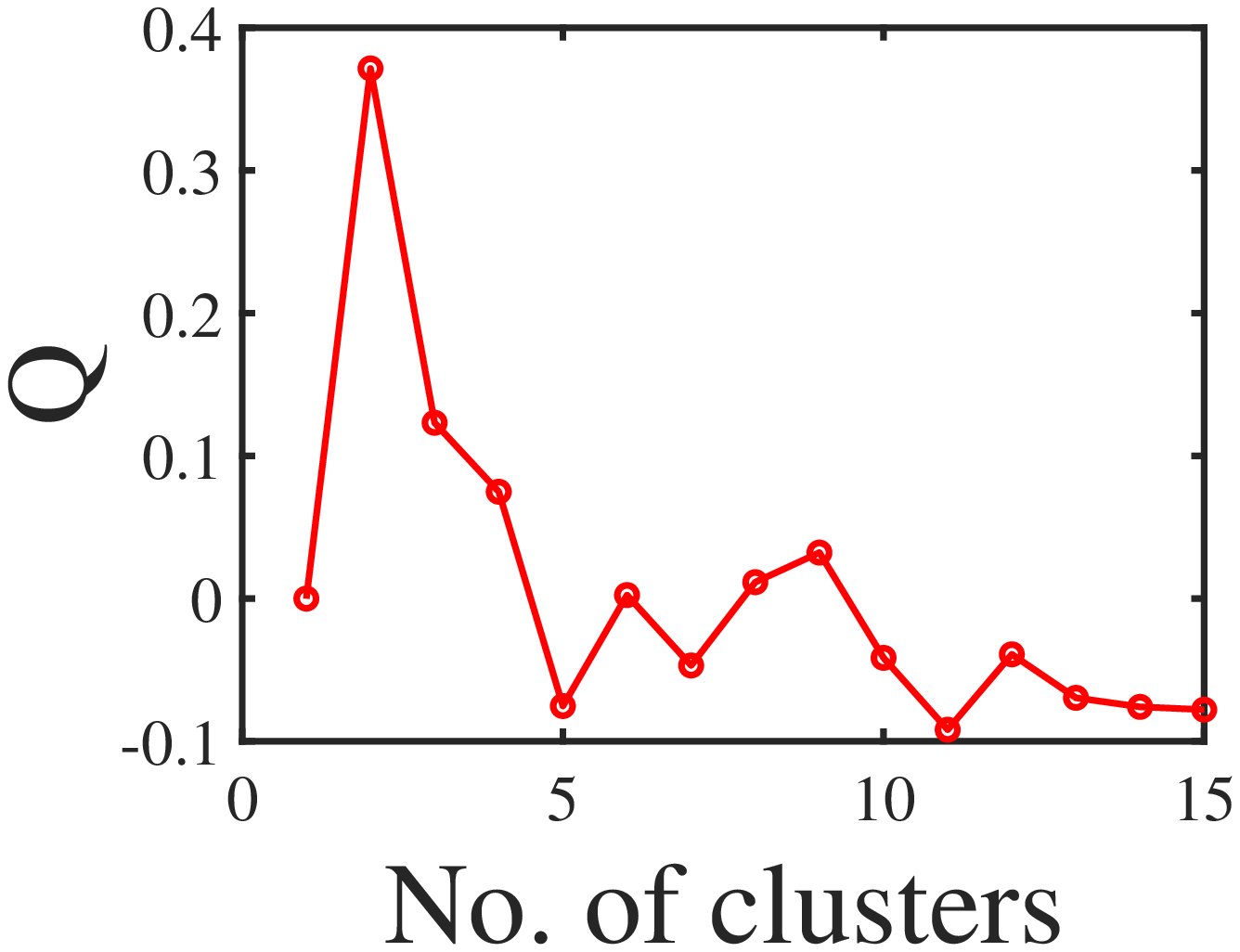}}
\subfigure[Political books]{\includegraphics[width=0.24\textwidth]{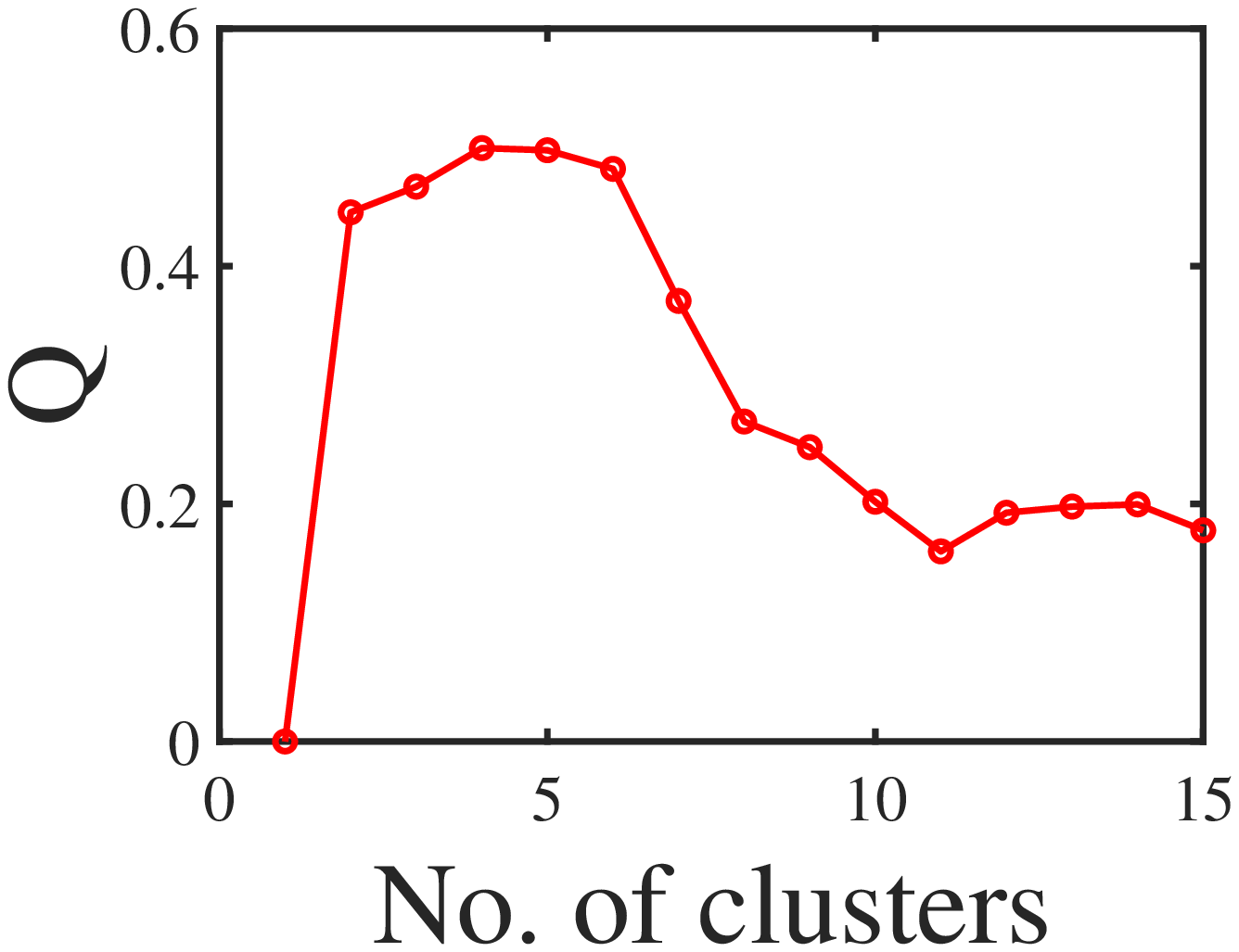}}
\subfigure[Political blogs]{\includegraphics[width=0.24\textwidth]{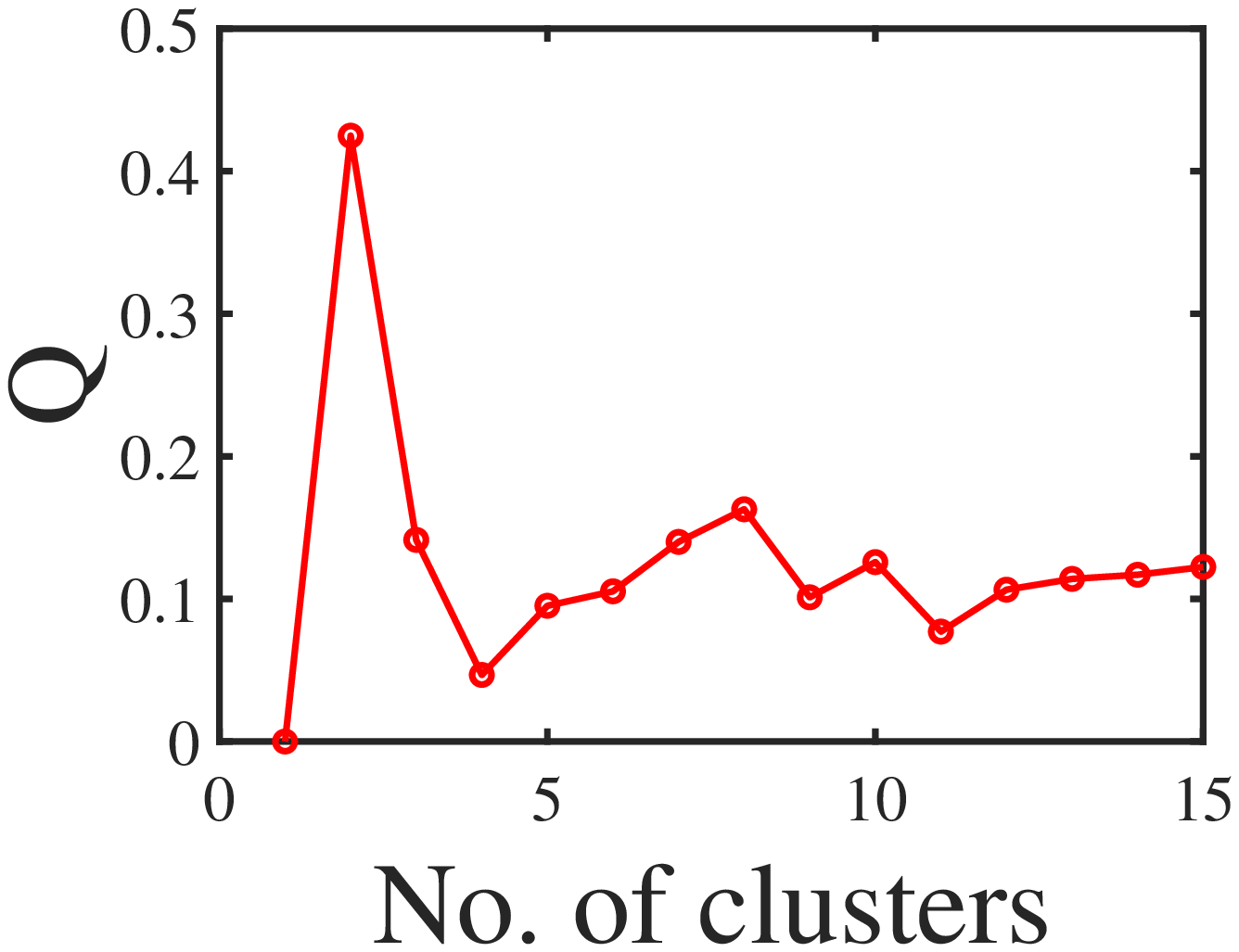}}
\caption{Weighted modularity $Q$ obtained from Equation (\ref{Modularity}) against the number of clusters by the nDFA algorithm for real-world networks considered in this paper.}
\label{Q} 
\end{figure}

\begin{table}[h!]
\footnotesize
	\centering
	\caption{Comparison of estimated K in real-world networks.}
	\label{realdata}
\begin{tabular}{ccccc|ccccccc}
\hline
\textbf{Dataset}&Source&$n$&$K$&Weighted?&nDFAwm&ME&NB&BHm&BHa&BHmc&BHac\\
\hline
Karate club (weighted)&\cite{zachary1977information}&34&2&Yes&2&2&4&4&4&4&4\\
Gahuku-Gama subtribes&\cite{read1954cultures}&16&3&Yes&3&N/A&1&1&12&N/A&13\\
Slovene Parliamentary Party network&\cite{ferligoj1996analysis}&10&2&Yes&2&2&N/A&N/A&N/A&N/A&N/A\\
Dolphins&\cite{dolphins0}&62&2,4&No&4&2&2&2&2&2&2\\
College football&\cite{football}&110&11&No&11&10&10&10&10&10&10\\
Karate club&\cite{zachary1977information}&34&2&No&2&34&2&2&2&2&2\\
Political books&\cite{newman2006finding}&105&3&No&4&2&3&3&4&4&4\\
Political blogs&\cite{Polblogs1}&1222&2&No&2&2&7&7&7&8&8\\
\hline
\end{tabular}
\end{table}

For real-world networks, we compare our nDFAwm with the modularity eigengap (ME) \cite{budel2020detecting}, NB \cite{le2022estimating}, BHm \cite{le2022estimating}, BHa \cite{le2022estimating}, BHmc \cite{le2022estimating}, and BHac \cite{le2022estimating}. For our nDFAwm, we take $K_{c}=n$. Figure \ref{Q} displays the weighted modularity from Equation (\ref{Modularity}) by the nDFA algorithm for different choices of the number of clusters and we can find the nDFAwm's estimated $K$ of the eight real-world networks from Figure \ref{Q} directly. Table \ref{realdata} shows the estimated number of clusters for these networks. For all networks except for the Political books network, our nDFAwm successfully determines the correct number of communities. For the ME method, it estimates the correct $K$ for Karate club (weighted), Slovene Parliamentary Party Network,  Dolphins, and Political blogs while it fails for the other four networks. For NB and BHm methods, they only estimate $K$ correctly for Dolphins, Karate club, and Political books. For BHa, BHmc, and BHac, they only estimate $K$ successfully for Dolphins and Karate club. In particular, the non-backtracking method and Bethe Hessian matrix-based methods proposed in \cite{le2022estimating} fail to estimate the number of communities for the three real-world weighted networks in Table \ref{realdata}. As a result, our nDFAwm outperforms its competitors in these real-world networks.
\section{Discussion}\label{Conclusion}
In this paper, we propose a method for determining the number of communities for weighted networks in DCDFM. We develop the method based on a combination of weighted modularity and a spectral clustering algorithm. This estimation method enables us to estimate the number of communities even in the case where there is only one community in a weighted network generated by different distributions under DCDFM. Through substantial computer-generated weighted networks from DCDFM and several real-world networks, the numerical results show that the estimation accuracy of our approach is better than its competitors and our method also works for signed networks.

There are some open questions. First, building a theoretical guarantee on the consistency of our estimator for the true number of clusters under DCDFM is an attractive and challenging task. Second, determining the exact condition under which estimating the number of clusters is possible under DCDFM is a challenging problem. Third, in this paper, we are mainly interested in DCDFM for non-overlapping networks, but the idea can be extended to overlapping weighted networks. Fourth, in this paper, we estimate the number of communities for weighted networks generated from DCDFM by Equation (\ref{SpectralModularity}) when we choose the method $\mathcal{M}$ as the spectral method nDFA. If we let $\mathcal{M}$ be algorithms developed in \cite{aicher2015learning, jog2015information,ahn2018hypergraph, palowitch2018significance,peixoto2018nonparametric,xu2020optimal,ng2021weighted} to fit their weighted stochastic blockmodels for weighted networks, we wonder that we can also estimate the number of communities for these weighted models through Equation (\ref{SpectralModularity}). We leave them for the future.
request.
\bibliography{refKDCDFM}
\end{document}